\documentclass[aps,pre,showpacs,twocolumn]{revtex4}
\usepackage{bm}
\usepackage{graphicx}
\bibstyle{approve.bib}
\usepackage{amssymb}
\usepackage{amsmath,empheq}
\usepackage{esint}
\usepackage{epstopdf}
\usepackage{color}
\usepackage{gensymb}
\usepackage{mathtools}
\usepackage{suffix}

\newcommand{\be}{\begin{equation}}
\newcommand{\ee}{\end{equation}}
\newcommand{\bea}{\begin{eqnarray}}
\newcommand{\eea}{\end{eqnarray}}

\newcommand{\br}{\mathbf{r}}

\newcommand{\bE}{\mathbf{E}}

\newcommand{\e}{\varepsilon}

\newcommand{\td}{\tilde{d}}
\newcommand{\ta}{\tilde{a}}
\newcommand{\lel}{\lambda_e}
\newcommand{\lb}{\lambda_b}
\newcommand{\lm}{L_-}
\newcommand{\lp}{L_+}

\begin{document}

\title{Controlling Polymer Capture and Translocation by Electrostatic Polymer-Pore Interactions}

\author{Sahin Buyukdagli$^{1}$\footnote{email:~\texttt{Buyukdagli@fen.bilkent.edu.tr}}  and 
T. Ala-Nissila$^{2,3}$\footnote{email:~\texttt{Tapio.Ala-Nissila@aalto.fi}}}
\affiliation{$^{1}$Department of Physics, Bilkent University, Ankara 06800, Turkey\\
$^{2}$Department of Applied Physics and COMP Center of Excellence, Aalto University School of Science, 
P.O. Box 11000, FI-00076 Aalto, Espoo, Finland\\
$^{3}$Departments of Mathematical Sciences and Physics, Loughborough University, Loughborough, 
Leicestershire LE11 3TU, United Kingdom}

\date{\small\it \today}

\begin{abstract}
Polymer translocation experiments typically involve anionic polyelectrolytes such as DNA molecules driven through 
negatively charged nanopores. Quantitative modelling of polymer capture to the nanopore followed by translocation
therefore necessitates the consideration of the electrostatic barrier resulting from like-charge polymer-pore interactions. To this end, in
this work we couple mean-field level electrohydrodynamic equations with the Smoluchowski formalism to characterize the interplay 
between the electrostatic barrier, the electrophoretic drift, and the electro-osmotic liquid flow. In particular, we find that due to distinct ion 
density regimes where the salt screening of the drift and barrier effects occur, there exists a characteristic salt concentration 
maximizing the probability of barrier-limited polymer capture into the pore. We also show that in the barrier-dominated regime, 
the polymer translocation time $\tau$ increases exponentially with the membrane charge 
and decays exponentially fast with the pore radius 
and the salt concentration.
These results suggest that the alteration of these parameters in the barrier-driven regime can be an 
efficient way to control the duration of the translocation process and facilitate more accurate measurements
of the ionic current signal in the pore.

\end{abstract}


\date{May 10, 2017}
\maketitle

\section{Introduction}

Biopolymer sequencing is of major relevance to various fields ranging from forensic sciences to biotechnology and gene therapy. 
In this context, nanopore-based sequencing approaches have been a central focus over the past two decades. 
Polymer translocation was initially conceptualised by using biological nanopores such as $\alpha$-Hemolysin 
channels of limited characteristics and undesirable fragility~\cite{e1,e2,e3,e4,e5,e6,e7,e8,e9}. 
Recent advancements in nanotechnology have significantly improved the reliability of the sequencing techniques. 
More precisely, the use of solid-state nanopores of various size and charge compositions now offers a wide range of 
functionalities that can allow to improve the resolution of the
method~\cite{e10,e11,e12,e13,e14,e15,e16,e17,e18,e19,e20,Tapsarev,e21,e22}. 
The technological progress requires development of theoretical models that can relate the tunable system parameters to 
experimentally observable quantities such as polymer capture rates, translocation times, and the ionic current blockade. 
Due to the high complexity of the polymer translocation process this constitutes a challenging task.

There are various factors that contribute to the complexity of the polymer translocation problem. The first difficulty stems from the non-equilibrium 
nature of the polymer capture and transport processes. Further, the entangled effect of different mechanisms on 
translocation such as electrostatic polymer-pore  and polymer-ion interactions, 
hydrodynamic polymer-solvent interactions, and conformational polymer fluctuations 
necessitates the consideration of these features on an equal footing. Thus, polymer translocation should be 
formulated within the framework of a beyond-equilibrium electrohydrodynamic theory which has not been
accomplished to date. 

Most models of polymer translocation dynamics to date are based on either 
coarse-grained computer simulations and theories that do not
explicitly take into account electrostatic effects, or short time scale
Molecular Dynamics (MD) simulations of atomistic polymer-pore models \cite{Tapsarev}. 
However, there are also theoretical attempts to consider some
specific aspects of electrostatics to translocation dynamics at the continuum level.
By coupling the mean-field (MF) Poisson-Boltzmann (PB) equation with the Stokes equation, 
Ghosal investigated the effect of salt on the DNA translocation velocity~\cite{the2,the3}. The influence of the 
polymer's self-energy on the unzipping of a DNA hairpin during translocation was studied by 
Zhang and Shklovskii in Ref.~\cite{the4}. Solving the linear PB equation together with the 
Smoluchowski equation, Wong and Muthukumar focused on the effect of the electro-osmotic flow on DNA 
capture outside the nanopore~\cite{the5}. A non-equilibrium theory of polymer transport through neutral pores was later 
developed by Muthukumar~\cite{the7,the15}. The polymer capture process with a detailed consideration of the 
polymer hydrodynamics was also modelled in Refs.~\cite{the8,the12,the13,the14}. Hatlo {\it et al.} investigated the effect of salt gradient on 
polymer capture~\cite{the10}. One of the central issues here is the reduction of the polymer's 
velocity upon its penetration into the pore in order to control the translocation process and readout of the ionic blockade current~\cite{e8}. MD simulations~\cite{the9} and correlation-corrected theories~\cite{the16} have 
shown that this goal can be achieved by the addition of polyvalent cations to the electrolyte solution.

Polymer translocation experiments are usually conducted with negatively charged polyelectrolytes such as 
DNA molecules translocating through silicon-based membrane nanopores carrying fixed negative charges on their wall~\cite{e19,Tapsarev}. 
The interaction between the pore and polymer charges is expected to result in an electrostatic barrier that opposes the polymer 
capture by the pore. To our knowledge, the effect of this barrier has not been taken into account by previous theories. 
Motivated by these points, in this work we develop a non-equilibrium polymer transport theory 
that treats on the same footing the electrostatic barrier, the electrophoretic drift, and the electroosmotic flow. 
In our model, we neglect conformational polymer fluctuations and treat the polyelectrolyte as a rigid charged cylinder.
Furthermore, we focus on the case of symmetric monovalent electrolytes and large pores where the 
PB formalism is known to be accurate~\cite{the16}. Therefore, we restrict ourselves to the 
MF formulation of electrostatic interactions. However, we note that our formalism is general enough for further extensions,  including electrostatic correlation effects that will be considered in future work.

Our polymer translocation model is developed in Section~\ref{for}. The formalism is based on the coupling of the 
Smoluchoswki equation with the PB and Stokes equations, and the force-balance relation for the polymer. 
In the inclusion of the electrostatic barrier, which is the main novelty of our work, we make use of a test-charge 
approach recently developed by one of us in Ref.~\cite{Buyuk2017}. By considering the steady-state regime of this 
electrohydrodynamically enhanced Smoluchowski formalism, we calculate the polymer translocation rate.  
The competition between the electrophoretic drift, the electroosmotic flow, and the electrostatic barrier is fully 
scrunitized in Section~\ref{res}. In the same section, we also investigate the effect of tunable system parameters on the 
polymer translocation time. Finally, we summarize our main results and discuss the 
approximations and potential extensions of our modeling.

\section{Polymer translocation model}
\label{for}
\begin{figure}
\includegraphics[width=1.\linewidth]{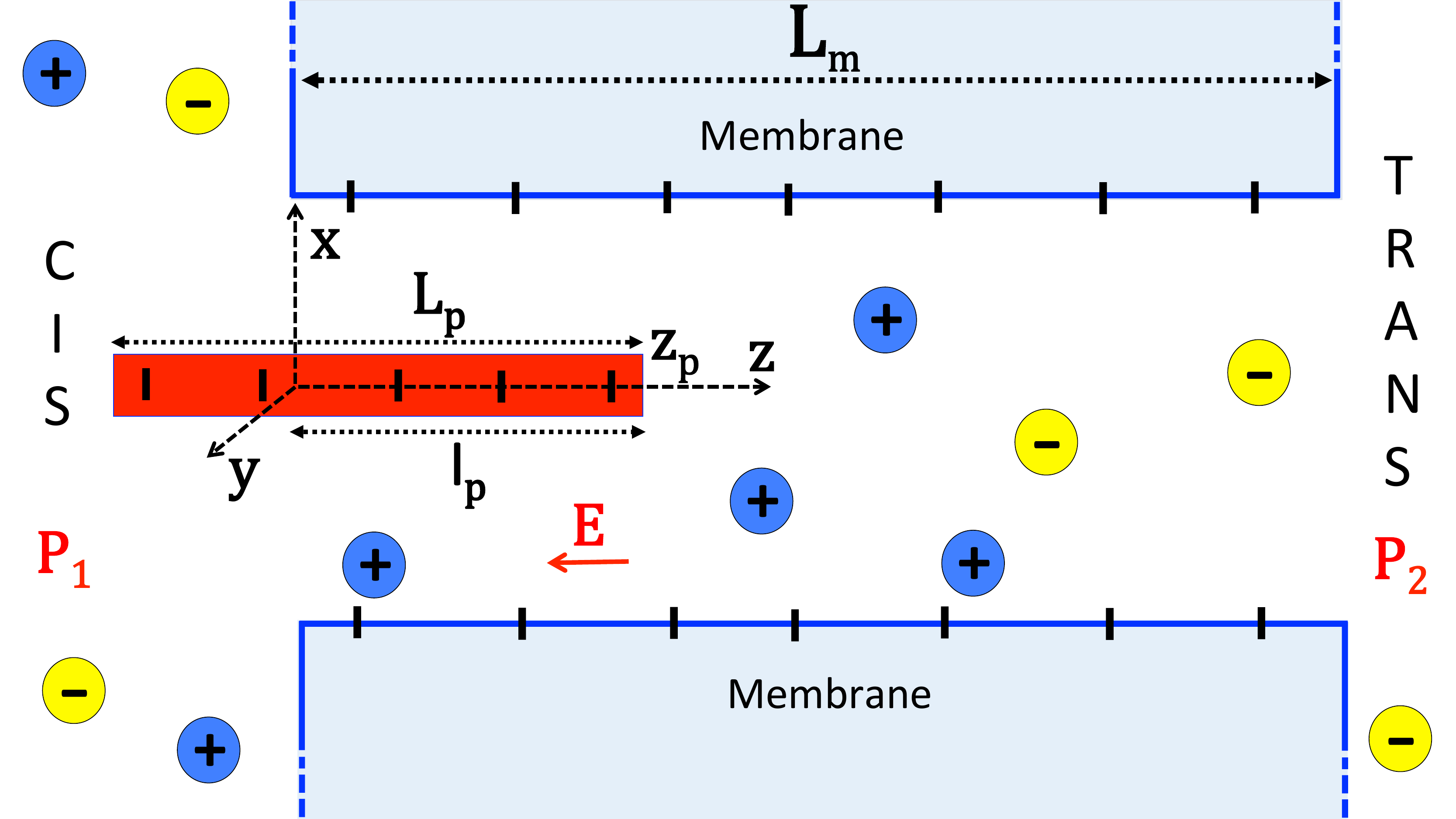}
\includegraphics[width=1.\linewidth]{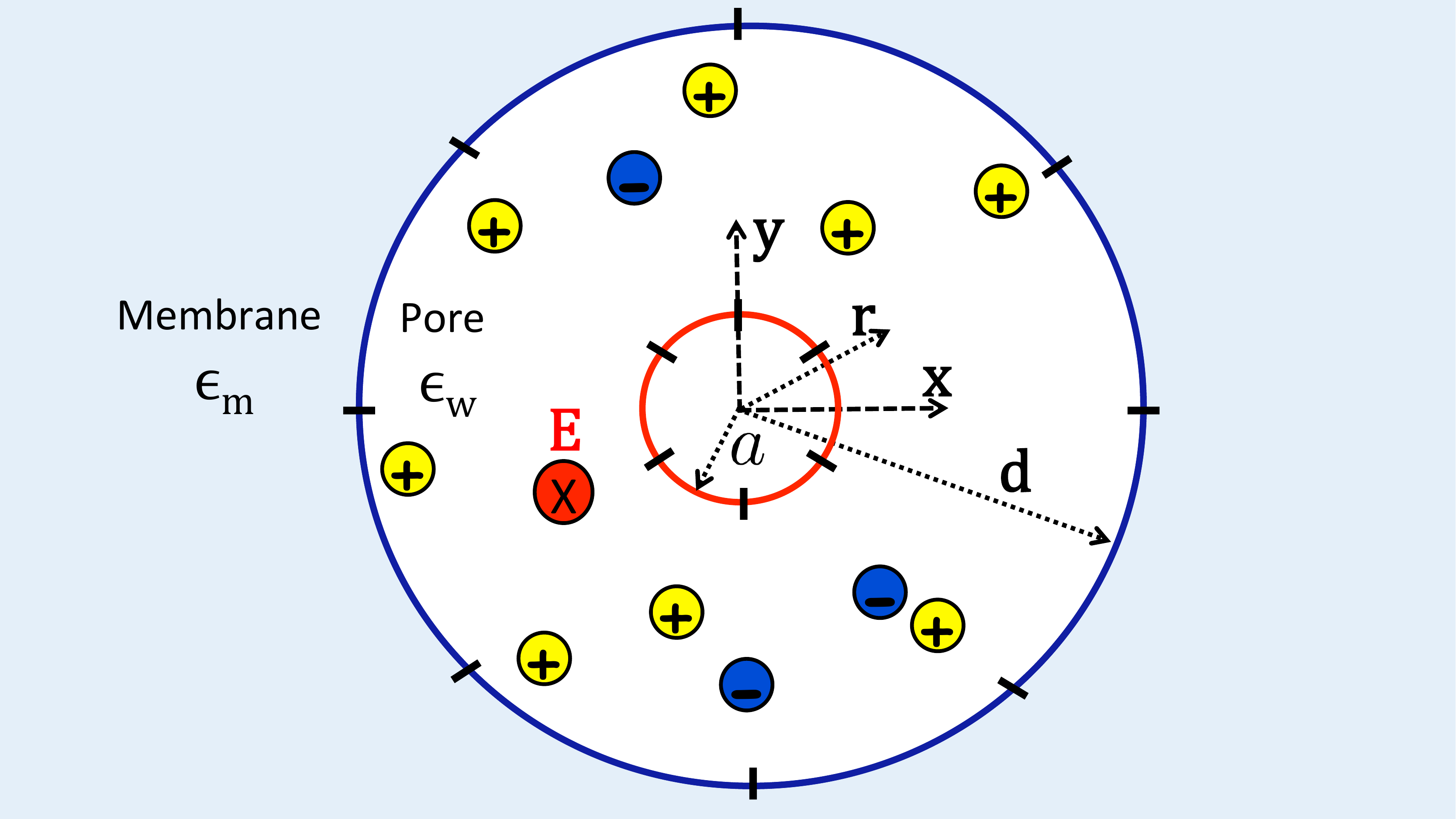}
\caption{(Color online) Schematic representation of the model of a translocating rigid polymer through the nanopore: 
side view (top panel) and top view (bottom panel). The cylindrical polymer has radius $a$, length $L_p$, and negative 
surface charge density $-\sigma_p$ with $\sigma_p>0$. The cylindrical nanopore has length $L_m$ (which may be
either longer or shorter than $L_p$),
radius $d$, and surface charge density $-\sigma_m$ with $\sigma_m>0$. The polymer portion in the pore has length $l_p$ 
with the right end located at $z=z_p$. The translocation takes place along the $z$ axis, with the external electric field 
$\bE=-E\hat{u}_z$ and pressure gradient $\Delta P=P_2-P_1$. }
\label{fig1}
\end{figure}

In this section, we derive the polymer translocation rates characterizing the barrier-limited capture of a polyelectrolyte and its transport 
through a charged pore confining an electrolyte solution.  The computation of the polymer translocation rate necessitates the 
steady-state solution of the Smoluchowski equation for the probability density of the polymer. To this end, in Sec.~\ref{sec1}, 
we derive an hydrodynamically enhanced Smoluchowski equation including the electrohydrodynamic properties of the 
translocating polymer and the surrounding charged liquid. The solution of this equation requires in turn the 
knowledge of the electrostatic potential in the pore as well as the electrostatic interaction energy of the polymer with the 
membrane. Based on MF level PB electrostatics, these features are derived in Sec.~\ref{elec}.

\subsection{Electrohydrodynamically augmented Smoluchowski equation}
\label{sec1}

The model of the charged polymer-pore system is depicted in Fig.~\ref{fig1}. The cylindrical nanopore has radius $d$ and length $L_m$. 
The pore wall carries negative fixed charges of density $-\sigma_m$ with $\sigma_m>0$.  The negatively charged polymer is a rigid 
cylinder of radius $a$, total length $L_p$, and uniform surface charge density $-\sigma_p$ with magnitude $\sigma_p>0$. 
The reservoir and the pore also contain a symmetric electrolyte composed of monovalent positive and negative charges with 
bulk concentration $\rho_{b}$.  We assume that the translocation takes place along the $z$ axis whose origin is located at the 
pore entrance. That is, we neglect off-axis polymer fluctuations. The reaction coordinate of the translocation is $z_p$,  
the position of the right end of the polymer. The length of the polymer portion located inside the pore will be denoted by $l_p$. In addition to the hydrodynamic drag force and the externally applied field $\bE=-E\hat{u}_z$ of magnitude $E$ along the negative $z$ axis, 
upon its penetration to the pore the polymer experiences an electrostatic barrier $V_p(z_p)$ resulting from its direct 
electrostatic interaction with the membrane. This electrostatic barrier will be derived in Sec.~\ref{grpol}.

The probability density of the polymer $c(z_p,t)$ solves the Smoluchowski equation that can be expressed as a continuity equation
\be
\label{a1}
\frac{\partial c(z_p,t)}{\partial t}=-\frac{\partial J(z_p,t)}{\partial z_p},
\ee
with the density current
\be
\label{a2}
J(z_p,t)=-D\frac{\partial c(z_p,t)}{\partial z_p}+c(z_p,t)v_p(z_p).
\ee
The first term on the r.h.s. of Eq.~(\ref{a2}) is the diffusive flux of entropic origin corresponding to Fick's law. 
The quantity $D$ stands for the translational diffusion coefficient of a cylindrical rigid polymer~\cite{cyl1,cyl2} given by
\be
\label{a2II}
D=\frac{k_B T \ln(L_p/2a)}{3\pi\eta L_p},
\ee
with the viscosity coefficient of water $\eta=8.91\times 10^{-4}\;\mathrm{Pa}\;\mathrm{s}$. We note that Eq.~(\ref{a2II}) is valid for $L_p\gg a$. The second term in the polymer current Eq.~(\ref{a2}) is the convective contribution from the polymer motion associated with external effects such as the applied field $E$, the hydrodynamic drag force on the polymer, and electrostatic polymer-pore interactions. By coupling the Stokes equation with the Poisson equation and the force balance relation, we derive next the corresponding polymer velocity $v_p(z_p)$.

\subsubsection{Computing the polymer velocity}

We assume that the convective liquid velocity is purely longitudinal and depends exclusively on the radial coordinate $r$. 
Therefore, the liquid velocity $u_c(r)$ solves the Stokes equation in the radial direction
\be\label{a2III}
\eta\nabla^2_r u_c(r)-eE\rho_c(r)+\frac{\Delta P}{L_m}=0,
\ee
where $e$ stands for the electron charge and $\rho_c(r)$ the ionic charge density. 
Here we combine the Stokes equation with the Poisson equation 
$\nabla^2_r \phi(r)+4\pi\ell_B\rho_c(r)=0$ for the average electrostatic potential $\phi(r)$ in the pore, 
where $\ell_B\approx7$ {\AA} is the Bjerrum length. This yields
\be
\label{a3}
\partial_rr\partial_ru_c(r)=-\mu_eE\partial_rr\partial_r\phi(r)-\frac{\Delta P}{\eta L_m}r,
\ee
where we have defined the electrophoretic mobility
\be\label{a3II}
\mu_e=\frac{\e_wk_BT}{e\eta},
\ee
where $\e_w=80$ the relative dielectric permittivity of water, $k_B$ the Boltzmann constant, and 
$T=300$ K the ambient temperature. Integrating Eq.~(\ref{a3}) twice we find
\be
\label{a4}
u_c(r)=-\mu_eE\phi(r)-\frac{\Delta P}{4\eta L_m}r^2+c_1\ln(r)+c_2.
\ee
In order to determine the integration constants $c_1$ and $c_2$, we impose a no-slip condition at the pore wall, i.e. $u_c(d)=0$. 
Next we account for the fact that at the polymer surface, the polymer and the liquid have the same velocity, 
$u_c(a)=v_p(z_p)$, where $z_p$ should be considered as an adiabatic variable. This yields the convective liquid velocity in the form
\bea\label{a5}
u_c(r)&=&-\mu_eE\left[\phi(r)-\xi_w\right]+\frac{\Delta P}{4\eta L_m}\left(d^2-r^2\right)\\
&&+\frac{\ln(d/r)}{\ln(d/a)} \left[ v_p(z_p)+\mu_eE(\xi_p-\xi_w)\right.\nonumber\\
&&\left.\hspace{1.5cm}-\frac{\Delta P}{4\eta L_m}\left(d^2-a^2\right) \right] ,\nonumber
\eea
where we introduced the polymer and pore surface potentials $\xi_p=\phi(a)$ and $\xi_w=\phi(d)$. These surface potentials will be 
explicitly calculated in Sec.~\ref{surp}.

At this point, we account for the force balance relation. This follows from the steady state regime of Newton's second law for the polymer, $F_e+F_d+F_b=0$, with the electrostatic force  on the DNA molecule $F_e=2\pi aL_p\sigma_p eE$, the hydrodynamic drag force $F_d=2\pi aL_p\eta u'_c(a)$, and the barrier-induced force $F_b=-V'_p(z_p)$. This yields
\be
\label{a6}
2\pi aL_p\left[\sigma_peE+\eta u'_c(a)\right]-\frac{\partial V_p(z_p)}{\partial z_p}=0.
\ee
Next, by using Eq.~(\ref{a5}) we eliminate the term $u_c'(a)$ in Eq.~(\ref{a6}). Accounting also for Gauss' law $\phi'(a)=4\pi\ell_B\sigma_p$, after some algebra the polymer velocity follows as
\be\label{a7}
v_p(z_p)=v_{\rm dr}+v_{\rm str}-\beta D_*\frac{\partial V_p(z_p)}{\partial z_p},
\ee
where $\beta = 1/(k_BT)$. In Eq.~(\ref{a7}), the first term is the drift velocity induced by the externally applied electric field $E$,
\be
\label{a8}
v_{\rm dr}=-\mu_e(\xi_p-\xi_w)E.
\ee
Since both the polymer and pore charges contribute to the surface potentials $\xi_p$ and $\xi_w$, Eq.~(\ref{a8}) includes 
both the electrophoresis and the effect of the electroosmotic liquid flow. The second term of Eq.~(\ref{a7}) is the 
velocity component associated with the streaming current,
\be
v_{\rm str}=\frac{\Delta P}{4\eta L_m}\left[d^2-a^2-2a^2\ln(d/a)\right].
\ee
Finally, the third term in Eq.~(\ref{a7}) corresponds to the effect of the barrier on the polymer velocity, 
with the effective diffusion coefficient in the pore
\be
\label{a9}
D_*=\frac{k_B T \ln(d/a)}{2\pi\eta L_p}.
\ee
We note that the effective diffusion coefficient $D_*$ is similar to the bulk value in 
Eq.~(\ref{a2II}), with the polymer length $L_p$ in the logarithm replaced by the pore radius $d$.

\subsubsection{Steady-state solution of the Smoluchowski equation}

In the steady-state regime of Eq.~(\ref{a1}) where $\partial_t c(z_p,t)=0$, the probability current is constant in time and uniform in the pore, i.e. $J(z_p,t)=J_0$. In this regime, plugging the velocity Eq.~(\ref{a7}) into Eq.~(\ref{a2}), the current becomes
\be
\label{a9II}
J_0=-D\frac{\partial c(z_p)}{\partial z_p}+c(z_p)\left[v_{\rm dr}+v_{\rm str}-\beta D_*\frac{\partial V_p(z_p)}{\partial z_p}\right].
\ee
Introducing the effective potential
\be
\label{a10}
U_p(z_p)=\frac{D_*}{D}V_p(z_p)-\frac{v_{\rm dr}+v_{\rm str}}{\beta D}z_p,
\ee
Eq.~(\ref{a9II}) can be expressed in the form
\be\label{a11}
e^{-\beta U_p(z_p)}\frac{\mathrm{d}}{\mathrm{d}z_p} \left[ c(z_p)e^{\beta U_p(z_p)} \right]=-\frac{J_0}{D}.
\ee
Finally, integrating Eq.~(\ref{a11}) the probability density of the polymer follows as
\be
\label{a12}
c(z_p)=\left[ C-\frac{J_0}{D}\int_0^{z_p}\mathrm{d}z\;e^{\beta U_p(z)} \right] e^{-\beta U_p(z_p)}.
\ee
The integration constants $C$ and $J_0$ in Eq.~(\ref{a12}) will be fixed by the boundary conditions. 
First, we assume that the polymer that leaves the pore is rapidly removed from the system. 
Thus, we impose an absorbing boundary condition at the point $z_p=L_m+L_p$, 
where the whole DNA molecule is located on the trans side, i.e. $c(L_m+L_p)=0$.  
The second condition follows from the polymer density at the pore entrance, $c(z_p=0)=c_{\rm out}$. 
Imposing these conditions to Eq.~(\ref{a12}) and considering that $U_p(0)=0$, the steady-state probability density becomes
\be\label{a15}
c(z_p)=c_{\rm out}\frac{\int_{z_p}^{L_m+L_p}\mathrm{d}z\;e^{\beta\left[U_p(z)-U_p(z_p)\right]}}{\int_{0}^{L_m+L_p}\mathrm{d}z\;e^{\beta U_p(z)}},
\ee
and the probability current reads $J_0=c_{\rm out}D/\int_0^{L_m+L_p}\mathrm{d}ze^{\beta U_p(z)}$. 
The polymer translocation rate is given by the ratio of the polymer current and the density at the pore entrance, i.e. $R_c=J_0/c_{\rm out}$, or
\be\label{a14}
R_c=\frac{D}{\int_0^{L_m+L_p}\mathrm{d}z\;e^{\beta U_p(z)}}.
\ee

\subsection{Electrostatic formalism}
\label{elec}

In this section we derive the electrostatic potential $\phi(\br)$ and the barrier $V_p(z_p)$ required for the computation of the drift 
and barrier-induced velocity components in Eq.~(\ref{a7}). In the present work, we will consider exclusively the case of monovalent 
electrolytes confined to large pores with radius $d>1$ nm where charge correlations are known to be negligible~\cite{the16}. 
Therefore, we will limit ourselves to the electrostatic MF formulation of the problem. However, it should be noted that the 
polymer transport formalism developed in Section~\ref{sec1} is not restricted to MF electrostatics and can be readily 
coupled with beyond-MF electrostatic equations. We will treat the corresponding charge correlation effects in a separate article.

\subsubsection{Computing the surface potentials and drift velocity}
\label{surp}

Here, we compute the drift velocity component $v_{\rm dr}$ of the polymer velocity Eq.~(\ref{a7}). According to Eq.~(\ref{a8}), 
this requires the derivation of the surface potentials $\xi_p=\phi(a)$ and $\xi_w=\phi(d)$. In the following calculation, we will neglect the longitudinal boundaries of the nanopore and the polymer. In order to compute the surface potentials, one has to solve the non-linear PB (NLPB) equation for a symmetric electrolyte,
\be\label{1}
\frac{1}{r}\partial_r\left[r\partial_r\phi(r)\right]+4\pi\ell_B\rho_c(r)=-4\pi\ell_B\left[\sigma_m(\br)+\sigma_p(\br)\right].
\ee
In Eq.~(\ref{1}), we introduced  the ion charge density function
\be\label{2}
\rho_c(r)=-2\rho_b\sinh\left[\phi(r)\right]\theta(r-a)\theta(d-r),
\ee
and the charge density of the polymer and the pore,
\bea\label{3}
\sigma_p(\br)&=&-\sigma_p\delta(r-a);\\
\label{4}
\sigma_m(\br)&=&-\sigma_m\delta(r-d).
\eea
Equation (\ref{1}) should be solved by imposing Gauss' law at the pore and polymer surface,
\be
\label{5}
\phi'(d^-)=-4\pi\ell_B\sigma_m\;;\hspace{5mm}\phi'(a^+)=4\pi\ell_B\sigma_p.
\ee

Equation (\ref{1}) cannot be solved analytically. Thus, we will solve it around the constant Donnan potential $\phi_d$ 
approximating the actual potential $\phi(r)$ in the pore. In order to determine the Donnan potential in Eq.~(\ref{1}),  we first
neglect the variations of the average potential and set $\phi(r)=\phi_d$. Integrating the resulting equation 
over the cross section of the pore, one gets
\be\label{6}
-2\rho_b\sinh(\phi_d)=\frac{2(\sigma_md+\sigma_pa)}{d^2-a^2},
\ee
whose inversion yields the Donnan potential
\be
\label{7}
\phi_d=-\ln\left(t+\sqrt{t^2+1}\right),
\ee
where we introduced the auxiliary parameter
\be
\label{8}
t=\frac{4}{\td^2-\ta^2}\left(\frac{\td}{s_m}+\frac{\ta}{s_p}\right).
\ee
In Eq.~(\ref{8}), we defined the adimensional radii $\td=\kappa_bd$ and $\ta=\kappa_ba$, where the bulk Debye-H\"{u}ckel 
parameter is given by $\kappa_b=\sqrt{8\pi\ell_B\rho_b}$. Furthermore, we introduced the parameters 
$s_m=\kappa_b\mu_m$ and $s_p=\kappa_b\mu_p$, where $\mu_m=1/(2\pi\ell_B\sigma_m)$ and $\mu_p=1/(2\pi\ell_B\sigma_p)$ stand for the 
Gouy-Chapman lengths associated with the membrane and polymer charges, respectively.

We can improve the Donnan approximation by accounting for the spatial variations of the potential in the pore. 
We express the average potential in the form
\be\label{9}
\phi(r)=\phi_d+\delta\phi(r).
\ee
Next, we insert Eq.~(\ref{9}) into Eq.~(\ref{1}) and Taylor expand the latter in terms of the correction term 
$\delta\phi(r)$. Using Eq.~(\ref{6}) and defining the Donnan screening parameter
\be\label{10}
\kappa_d=\sqrt{8\pi\ell_B\rho_b\cosh(\phi_d)}=\kappa_b\left(1+t^2\right)^{1/4},
\ee
one gets the differential equation 
\be\label{11}
\left(r^{-1}\partial_rr\partial_r-\kappa_d^2\right)\delta\phi(r)=-\frac{8\pi\ell_B}{d^2-a^2}(\sigma_md+\sigma_pa).
\ee
The solution to this linear differential equation satisfying the boundary conditions~(\ref{5}) reads
\bea
\label{12}
\delta\phi(r)&=&\frac{8\pi\ell_B}{\kappa_d^2}\frac{\sigma_md+\sigma_pa}{d^2-a^2}\\
&&+\frac{4\pi\ell_B}{\kappa_d}\frac{T_1\mathrm{I}_0(\kappa_dr)+T_2\mathrm{K}_0(\kappa_dr)}{\mathrm{I}_1(\kappa_da)\mathrm{K}_1(\kappa_dd)-\mathrm{K}_1(\kappa_da)\mathrm{I}_1(\kappa_dd)}\nonumber,
\eea
where we introduced the auxiliary parameters
\bea
\label{13}
T_1&=&\sigma_m\mathrm{K}_1(\kappa_da)+\sigma_p\mathrm{K}_1(\kappa_dd);\\
\label{14}
T_2&=&\sigma_m\mathrm{I}_1(\kappa_da)+\sigma_p\mathrm{I}_1(\kappa_dd).
\eea
In Eq.~(\ref{12}), we used the modified Bessel functions $\mathrm{I}_m(x)$ and $\mathrm{K}_m(x)$~\cite{math}. 
Using Eq.~(\ref{9}), the drift velocity~(\ref{a8}) can be expressed in terms of Eq.~(\ref{12}) as
\be\label{15}
v_{\rm dr}=-\mu_e\left[\delta\phi(a)-\delta\phi(d)\right]E.
\ee
In Sec.~\ref{velpr}, the accuracy of the improved Donnan approximation will be tested by comparing the drift velocity of
Eq.~(\ref{15}) with the result obtained from the numerical solution of the NLPB in Eq.~(\ref{1}) (see Fig. \ref{fig2}).

\subsubsection{Computing the electrostatic barrier}
\label{grpol}

In this subsection we calculate the electrostatic barrier experienced by the DNA inside the pore. In our model, the barrier 
$V_p(z_p)$ is induced by the electrostatic coupling between the DNA charges and the fixed charges on the nanopore wall. 
Thus, in the calculation of this barrier, we will neglect the electrostatic potential outside the pore and take into account only the 
polymer portion of length $l_p$ located in the pore. As translocation experiments cover a wide range of polymer and pore 
sizes, the total polymer length $L_p$ can be shorter or longer than the pore length $L_m$. In order to generalize the 
formulation of the problem to both situations, we introduce the auxiliary lengths
 \be
 \label{lmp}
 L_-=\mathrm{min}(L_m,L_p)\;;\hspace{5mm} L_+=\mathrm{max}(L_m,L_p).
 \ee
Hence, the barrier $V_p(z_p)$ can be expressed in terms of the electrostatic grand potential $\Omega_{\rm mf}(l_p)$ of the 
polymer portion in the pore as
\bea\label{a3III}
V_p(z_p)&=&\Omega_{\rm mf}(l_p=z_p)\theta(L_--z_p)\\
&&+\Omega_{\rm mf}(l_p=L_-)\theta(z_p-L_-)\theta(L_+-z_p)\nonumber\\
&&+\Omega_{\rm mf}(l_p=L_p+L_m-z_p)\theta(z_p-L_+).\nonumber
\eea
The first, second, and third terms of Eq.~(\ref{a3III}) correspond respectively to the polymer capture regime, the translocation 
at constant length $l_p=L_-$,  and the exit regime. 

In the MF limit of the test charge approach developed in Ref.~\cite{Buyuk2017}, the polymer grand potential reads
\be
\label{16}
\beta\Omega_{\rm mf}=\int\mathrm{d}\br\sigma_p(\br)\phi_m(\br),
\ee
In Eq.~(\ref{16}), $\phi_m(\br)$ is the average potential induced exclusively by the fixed charges on the membrane wall. 
Thus, this potential solves the PB Eq.~(\ref{1}) without the polymer charge density. 
Consequently, the potential $\phi_m(r)$ can be obtained from Eq.~(\ref{9}) by setting $\sigma_p=0$. This yields
\be\label{17}
\phi_m(r)=\phi_{\rm md}+\delta\phi_m(r),
\ee
with the Donnan potential $\phi_{\rm md}$ associated only with the pore charges
\be
\label{18}
\phi_{\rm md}=-\ln\left(t_m+\sqrt{t_m^2+1}\right),
\ee
where
\be
\label{19}
t_m=\frac{4\td s_m^{-1}}{\td^2-\ta^2}.
\ee
In Eq.~(\ref{17}), the potential correction $\delta\phi_m(r)$ follows from Eq.~(\ref{12}) in the form
\bea
\label{20}
\delta\phi_m(r)&=&\frac{8\pi\ell_B}{\kappa_m^2}\frac{\sigma_md}{d^2-a^2}\\
&&+\frac{4\pi\ell_B\sigma_m}{\kappa_m}\frac{\mathrm{K}_1(\kappa_{m}a)\mathrm{I}_0(\kappa_mr)+\mathrm{I}_1(\kappa_ma)\mathrm{K}_0(\kappa_mr)}{\mathrm{I}_1(\kappa_ma)\mathrm{K}_1(\kappa_md)-\mathrm{K}_1(\kappa_ma)\mathrm{I}_1(\kappa_md)},\nonumber
\eea
where we introduced the screening parameter associated with the charged pore only,
\be\label{21}
\kappa_m=\kappa_b\left(1+t_m^2\right)^{1/4}.
\ee

For the evaluation of the polymer grand potential~(\ref{16}), we will include into the polymer charge density Eq.~(\ref{3}) the 
length of the polymer portion located in the pore,
\be
\label{22}
\sigma_p(\br)=-\sigma_p\delta(r-a)\theta(z_p-z)\theta(z-z_p+l_p).
\ee
The MF grand potential~(\ref{16}) then becomes
\be\label{23}
\beta\Omega_{\rm mf}(l_p)=-2\pi al_p\sigma_p\phi_m(a).
\ee
Finally, substituting Eq.~(\ref{23}) into Eq.~(\ref{a3III}), the electrostatic barrier experienced by the polymer takes the form
\be\label{24}
\beta V_p(z_p)=-2\pi a\sigma_p\phi_m(a)\Theta(z_p),
\ee
where we introduced the piecewise function
\bea
\label{24II}
\Theta(z_p)&=&z_p\theta(L_--z_p)+L_-\theta(z_p-L_-)\theta(L_+-z_p)\nonumber\\
&&+(L_p+L_m-z_p)\theta(z_p-L_+).
\eea

\section{Results}
\label{res}  

Based on the drift velocity Eq.~(\ref{15}) and electrostatic barrier Eq.~(\ref{24}), 
we derive here the polymer velocity, translocation rates, and translocation time. In the rest of the article, 
we will consider the case of a vanishing pressure gradient $\Delta P=0$ which yields a vanishing streaming velocity 
$v_{\rm str}=0$ in Eq.~(\ref{a10}). We also note that unless otherwise stated, all results will be obtained from the improved 
Donnan approach of Eqs.~(\ref{15}) and~(\ref{24}).

\subsection{Polymer potential and velocity profile}
\label{velpr}

In order to derive the potential $U_p(z_p)$, we introduce the characteristic inverse lengths $\lel$ and $\lb$ associated 
respectively with the drift motion and the barrier,
\bea
\label{25}
\lel&=&\frac{\mu_e}{D}\left[\delta\phi(d)-\delta\phi(a)\right]E;\\
\label{25II}
\lb&=&-2\pi a\sigma_p\phi_m(a)\frac{D_*}{D}.
\eea
Injecting the drift velocity Eq.~(\ref{15}) and the barrier Eq.~(\ref{24}) into Eq.~(\ref{a10}), the effective potential becomes
\be
\label{26}
\beta U_p(z_p)=-\lel z_p+\lb\Theta(z_p),
\ee
where the piecewise function $\Theta(z_p)$ is defined in Eq.~(\ref{24II}). We derive next the polymer velocity $v_p(z_p)$ of Eq.~(\ref{a7}). 
According to Eqs.~(\ref{a7}) and~(\ref{a10}), the polymer velocity is related to the effective potential~(\ref{26}) by 
$v_p(z_p)=-\beta DU'_p(z_p)$. This yields the piecewise velocity profile
\bea
\label{27}
v_p(z_p)&=&\left(v_{\rm dr}-v_b\right)\theta(\lm-z_p)\nonumber\\
&&+v_{\rm dr}\theta(z_p-\lm)\theta(\lp-z_p)\nonumber\\
&&+\left(v_{\rm dr}+v_b\right)\theta(z_p-\lp),
\eea
where the drift and barrier-induced velocity components are respectively
\be
\label{27II}
v_{\rm dr}=D\lambda_e\;;\hspace{5mm}v_b=D\lambda_b.
\ee

\begin{figure}
\includegraphics[width=1.1\linewidth]{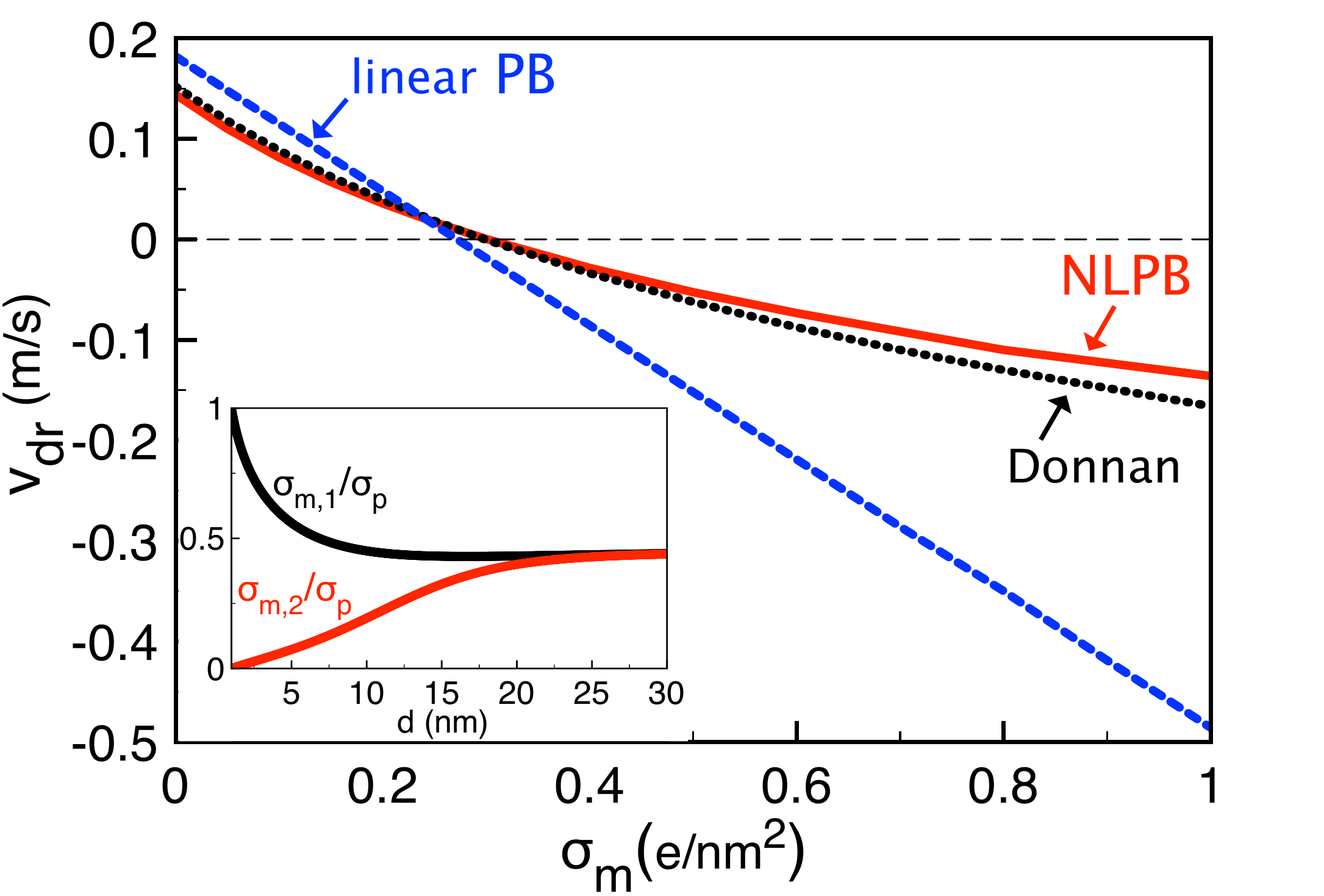}
\caption{(Color online) Main plot: Drift velocity component $v_{\rm dr}=-\mu_e\left[\phi(a)-\phi(d)\right]E$ 
versus the membrane charge $\sigma_m$ obtained from the numerical solution of the non-linear PB (NLPB) Eq.~(\ref{1}) (red), the 
Donnan approximation of Eq.~(\ref{15}) (black), and the solution of the linearized PB Eq.~(\ref{lin1}) (blue). 
The bulk salt concentration is $\rho_b=0.01$ M. The polymer charge is $\sigma_p=0.4$ $\mbox{e/nm}^2$ and radius $a=1$ nm. 
The pore has radius $d=3$ nm and length $L_m=34$ nm. The electric field is $E=\Delta V/L_m$ with the external voltage $\Delta V=120$ mV. 
The inset displays the critical membrane charges of Eqs.~(\ref{sigm1}) (black)  and~(\ref{sigm2}) (red) against the pore size.}
\label{fig2}
\end{figure}

\subsubsection{Drift velocity reversal}

The main plot of Fig.~\ref{fig2} displays the drift velocity component $v_{\rm dr}$ against the membrane charge $\sigma_m$. 
The red curve is the exact MF result obtained from the numerical solution of the PB Eq.~(\ref{1}). One notes that the 
Donnan approximation Eq.~(\ref{15}) (black curve) is significantly more accurate than the result obtained from the standard solution 
of the linear PB equation (blue curve),
\be
\label{lin1}
v_{\rm dr}=\frac{4\pi\ell_B\mu_eE}{g\kappa_b}(f_p\sigma_p-f_m\sigma_m).
\ee
In Eq.~(\ref{lin1}),  we introduced the geometric coefficients
\bea
\label{fp}
f_p&=&\mathrm{K}_1(\td)\mathrm{I}_0(\ta)+\mathrm{I}_1(\td)\mathrm{K}_0(\ta)-\td^{-1};\\
f_m&=&\mathrm{K}_1(\ta)\mathrm{I}_0(\td)+\mathrm{I}_1(\ta)\mathrm{K}_0(\td)-\ta^{-1};\\
g&=&\mathrm{I}_1(\td)\mathrm{K}_1(\ta)-\mathrm{I}_1(\ta)\mathrm{K}_1(\td),
\eea
with $\ta=\kappa_ba$ and $\td=\kappa_bd$. Equation (\ref{lin1}) can be derived alternatively from the 
Taylor expansion of Eq.~(\ref{15}) in terms of the charge densities $\sigma_m$ and $\sigma_p$. The main point in 
Fig.~\ref{fig2} is the change of the sign of the velocity from positive to negative with increasing membrane charge. 
This stems from the counterion attraction by the charged pore, which results in an electroosmotic flow moving 
parallel with the field~\cite{the3}.  At large membrane charges $\sigma_m\gtrsim0.3$, the hydrodynamic drag 
exerted by this flow on the polymer dominates the electric force induced directly by the field $E$ on the polymer charges. 
This reverses the direction of the drift velocity component $v_{\rm dr}$ which becomes negative. 

According to Eq.~(\ref{lin1}), the reversal of the drift velocity occurs at membrane charge densities 
$\sigma_m\geq\sigma_{m,1}$ with the threshold charge $\sigma_{m,1}$ given by
\be
\label{sigm1}
\frac{\sigma_{m,1}}{\sigma_p}=\frac{f_p}{f_m}.
\ee
Equation (\ref{sigm1}) is plotted versus the pore size in the inset of Fig.~\ref{fig2}. First, one notes that $\sigma_{m,1}<\sigma_p$ 
for any pore size. Then, at large pore radii $\td\gg1$ the characteristic charge $\sigma_{m,1}$ converges to the saturation value 
$\sigma_{m,1}\approx\sigma_p\mathrm{K}_0(\ta)/\mathrm{K}_1(\ta)$. With decreasing polymer radius $a$, 
this saturation value is lowered according to the relation $\sigma_{m,1}/\sigma_p\approx-\ta\ln\ta$ for $\ta\ll1$.

\begin{figure}
\includegraphics[width=1.\linewidth]{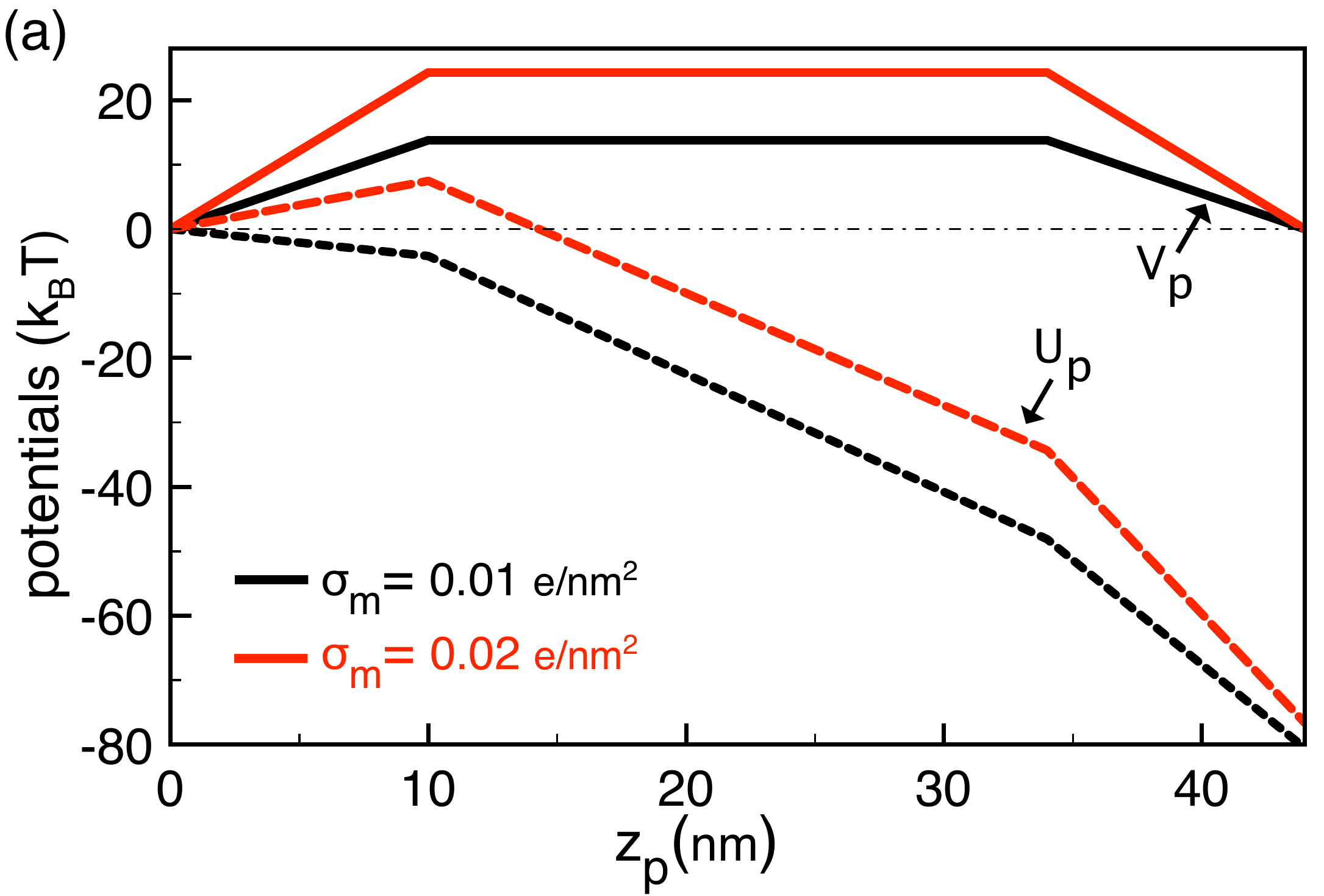}
\includegraphics[width=1.\linewidth]{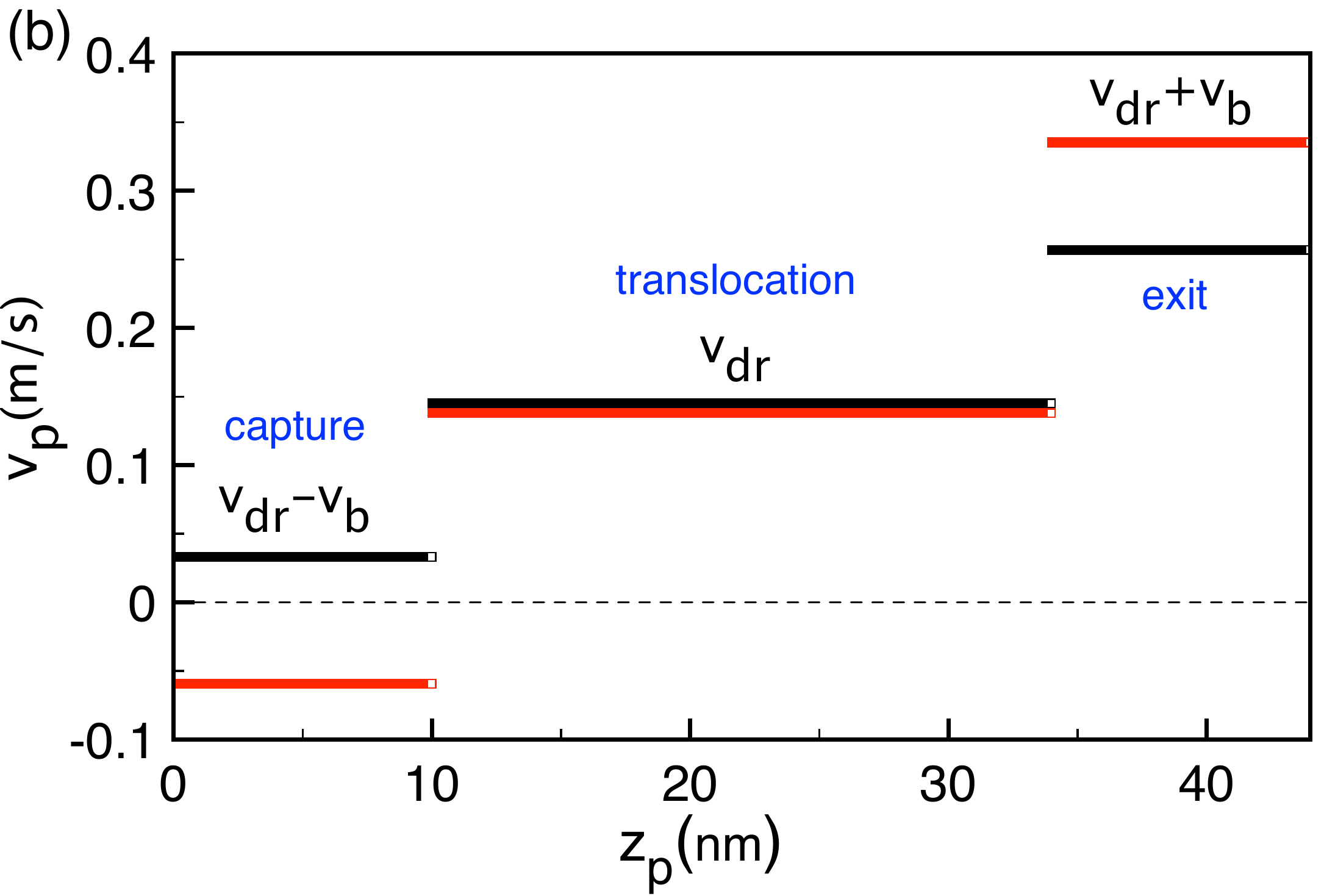}
\caption{(Color online) (a) Electrostatic barrier Eq.~(\ref{24}) (solid curves) and polymer potential Eq.~(\ref{26}) (dashed curves) 
versus the polymer position. (b) Polymer velocity profile Eq.~(\ref{27}). In (a) and (b), the membrane charge is 
$\sigma_m=0.01$ $\mathrm{e/nm}^2$ (black curves) and $0.02$ $\mathrm{e/nm}^2$ (red curves). 
The polymer and pore lengths are $L_p=\lm=10$ nm and $L_m=\lp=34$ nm. The remaining parameters are the same as in Fig.~\ref{fig2}.
See text for details.}
\label{fig3}
\end{figure}

\subsubsection{Influence of electrostatic barrier on polymer velocity}

We investigate next the influence of the membrane charge $\sigma_m$ on the net polymer velocity $v_p(z_p)$. To this end, 
in Figs.~\ref{fig3}(a) and (b) we plot the electrostatic barrier Eq.~(\ref{24}), the polymer potential Eq.~(\ref{26}), 
and the velocity profile Eq.~(\ref{27}) at two different membrane charges given in the legend. 
Figures \ref{fig3}(a) and (b) should be interpreted together.  We focus first on the membrane charge value 
$\sigma_m=0.01$ $\mbox{e/nm}^2$ (black curves). During the polymer capture regime $z_p\leq\lm$,  the barrier 
$V_p(z_p)$ that rises linearly with the position $z_p$ lowers the polymer velocity to $v_p(z_p)=v_{\rm dr}-v_b=D(\lel-\lb)$. 
In the translocation regime $\lm\leq z_p\leq\lp$ where the length of the polymer portion is constant in the pore, 
$l_p=\lm=\mbox{min}(L_p,L_m)$, the barrier $V_p(z_p)$ is constant and the polymer velocity is purely drift imposed, 
i.e. $v_p(z_p)=v_{\rm dr}=D\lel$.  As the polymer gets into the exit regime $z_p>\lp$ where the potential $V_p(z_p)$ is 
downhill, the polymer velocity is enhanced to the value $v(z_p)=v_{\rm dr}+v_b=D(\lb+\lel)$. 

Figure \ref{fig3}(a) shows that the external field $E$ drops the net potential $U_p(z_p)$ experienced by the polymer below the barrier 
$V_p(z_p)$. At the membrane charge $\sigma_m=0.01$ $\mbox{e/nm}^2$ corresponding to the drift-dominated regime with 
$\lel>\lb$ (black curves), the potential $U_p(z_p)$ is downhill for $z_p\leq\lm$ and the capture velocity in Fig.~\ref{fig3}(b)  
positive, $v_p=v_{\rm dr}-v_b=D(\lel-\lb)>0$. Rising the membrane charge to $\sigma_m=0.02$ $\mbox{e/nm}^2$ 
where one gets into the barrier-dominated regime with $\lb>\lel$ (red curves), the barrier $V_p(z_p)$ is enhanced and the potential 
$U_p(z_p)$ turns from downhill to uphill for $z_p\leq\lm$. Consequently, at the pore entrance, the polymer velocity changes its direction 
and becomes negative, $v_p=v_{\rm dr}-v_b<0$. Thus, at this membrane charge value and beyond, 
the polymer is likely to be rejected from the pore. The transition from drift to barrier-dominated regime is investigated in 
Sec.~\ref{tranrate} in terms of the polymer translocation rate.

\subsection{Polymer capture and translocation rates}
\label{tranrate}

Here, we calculate the polymer translocation rate. Evaluating the integral in Eq.~(\ref{a14}) with the potential function~(\ref{26}), the polymer translocation rate follows as
\be\label{28}
R_c=\frac{R_1R_2R_3}{R_1R_2+R_2R_3+R_1R_3},
\ee
where the characteristic rates for barrier-limited polymer capture, translocation at constant length, and exit regimes are respectively given by
\bea
\label{29I}
R_1&=&\frac{D(\lel-\lb)}{1-e^{-L_-(\lel-\lb)}};\\
\label{29II}
R_2&=&\frac{D\lel e^{-\lb\lm}}{e^{-\lel\lm}-e^{-\lel\lp}};\\
\label{29III}
R_3&=&\frac{D(\lel+\lb)e^{-\lb(L_p+L_m)}}{e^{-(\lel+\lb)L_+}-e^{-(\lel+\lb)(L_p+L_m)}}.
\eea
Substituting Eqs.~(\ref{29I})-(\ref{29III}) into Eq.~(\ref{28}), we finally get
\begin{widetext}
\be\label{30}
R_c=\frac{D\lel(\lel^2-\lb^2)e^{\lel(L_p+L_m)}}{(\lel+\lb)e^{\lel\lp}\left[\lel e^{\lel\lm}-\lb e^{\lb\lm}\right]-(\lel-\lb)\left[\lel+\lb e^{(\lel+\lb)\lm}\right]}.
\ee
\end{widetext}
In the case of a neutral pore and vanishing external field $E=0$ where $\lel=\lb=0$, the translocation rate takes the simple diffusive form 
$R_c=D/(L_m+L_p)$. Next, we investigate the dependence of the translocation rate on the membrane charge $\sigma_m$ and pore radius $d$.

\subsubsection{Membrane charge $\sigma_m$ and pore radius $d$}

\begin{figure}
\includegraphics[width=1.\linewidth]{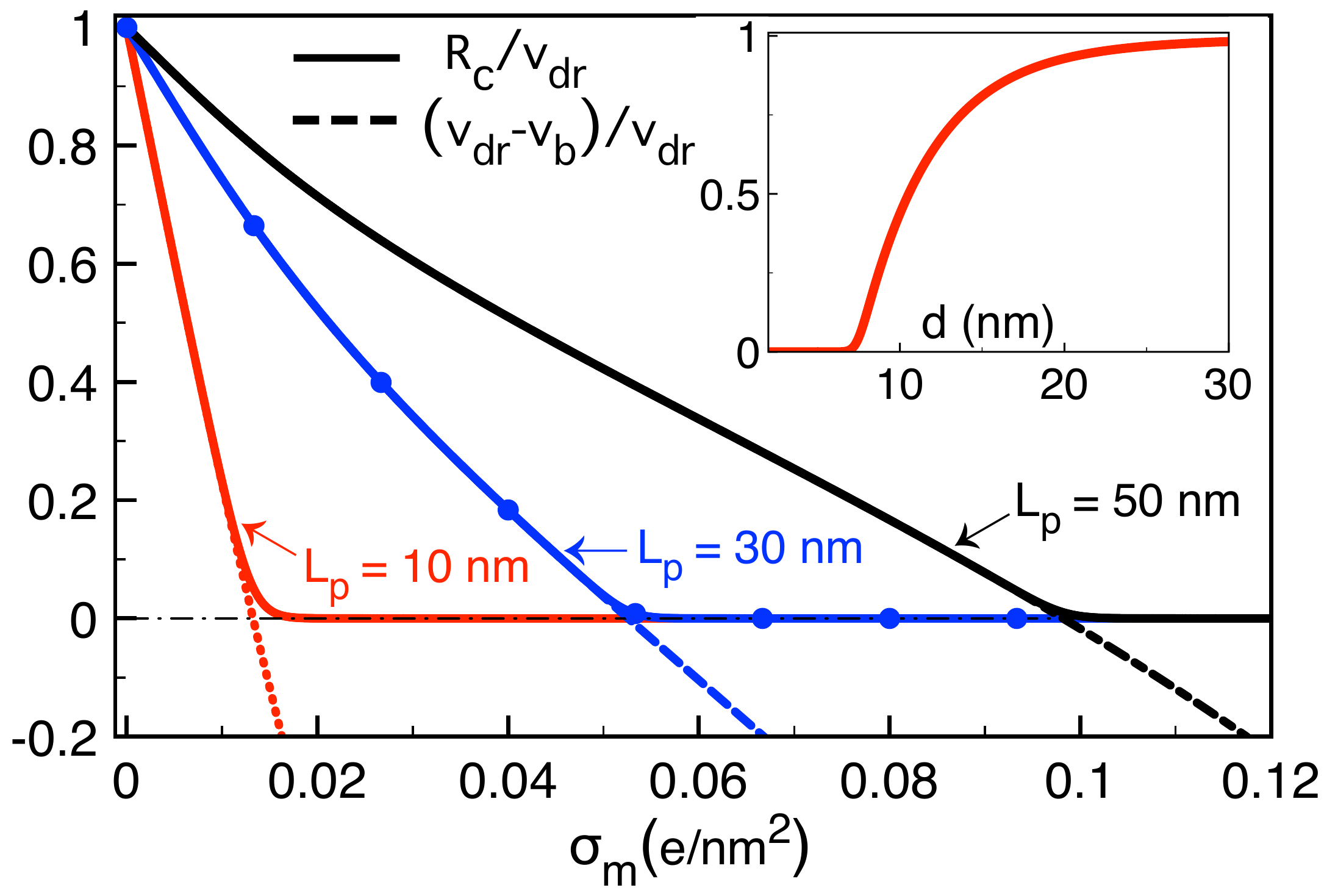}
\caption{(Color online) Polymer translocation rate $R_c$ (solid curves) and polymer capture velocity $v_{\rm dr}-v_b=D(\lel-\lb)$ 
(dashed curves) rescaled by the drift velocity $v_{\rm dr}$ against the membrane charge $\sigma_m$. The polymer lengths are 
$L_p=10$ nm (red curves), $L_p=30$ nm (blue curves), and $L_p=50$ nm (black curves). The inset displays the rescaled translocation 
rate versus the pore radius at the membrane charge $\sigma_m=0.05$ $\mbox{e/nm}^2$.  The dots in the main plot at $L_p=30$ nm 
correspond to the barrier-limited polymer capture rate $R_1$. The remaining parameters are the same as in Fig.~\ref{fig2}.}
\label{fig4}
\end{figure}

In Fig.~\ref{fig4}, we plot the translocation rate (solid curves) and the capture velocity $v_{\rm dr}-v_b$ (dashed lines) rescaled 
by the drift velocity $v_{\rm dr}$ against the membrane charge $\sigma_m$ at different polymer lengths $L_p$. 
We note that in the limit of a neutral pore $\sigma_m=0$, all curves converge to $R_c/v_{\rm dr}=1$. 
In this limit where the barrier vanishes ($V_p(z_p)=0$ and $\lambda_b=0$), the translocation rate~(\ref{30}) becomes
\be
\label{30II}
R_c=\frac{D\lel}{1-e^{-(L_p+L_m)\lel}}\approx v_{\rm dr}.
\ee
Thus, polymer transport through neutral pores is purely electrophoretic.

For the case of charged membranes Fig.~\ref{fig4} shows that in the drift-driven regime with $\lb<\lel$ or $\sigma_m<\sigma_{m,2}$ 
where the characteristic charge $\sigma_{m,2}$ will be calculated below, the translocation rate drops linearly with increasing membrane 
charge. In the subsequent barrier-dominated regime $\lb>\lel$ or $\sigma_m>\sigma_{m,2}$, the translocation rate decays exponentially. 

We investigate first the drift-dominated regime $\sigma_m<\sigma_{m,2}$. We note that the total translocation rate 
Eq.~(\ref{30}) can be very accurately approximated by the barrier-limited capture rate of Eq.~(\ref{29I}), i.e. $R_c\approx R_1$ 
(compare the blue curve and dots in Fig.~\ref{fig4}).  Thus, for $\lel>\lb$, the behaviour of the translocation rate follows from Eq.~(\ref{29I}) as 
\be
\label{as1}
R_c\approx D\left(\lel-\lb\right)\left[1+e^{-\lm\left(\lel-\lb\right)}\right]\approx v_{\rm dr}-v_b,
\ee
which explains the superposition of the velocity and translocation rate curves. We now note that in the linear PB approximation, 
the barrier-induced velocity component in Eq.~(\ref{27II}) takes the simple form
\be
\label{lin2}
v_b=\frac{4\pi\ell_B\ln(d/a)}{g\eta\beta L_p\kappa_b^2}\sigma_p\sigma_m.
\ee
Substituting the velocity components~(\ref{lin1}) and~(\ref{lin2}) into Eq.~(\ref{as1}), 
we get a closed-form expression for the translocation rate in the drift-dominated regime as
\be
\label{as1II}
R_c\approx\frac{4\pi\ell_B}{g\kappa_b}\left[\mu_eE(f_p\sigma_p-f_m\sigma_m)-\frac{\ln(d/a)}{\eta\beta\kappa_bL_p}\sigma_p\sigma_m\right].
\ee
The linear dependence of Eq.~(\ref{as1II}) on the membrane charge $\sigma_m$ explains the linear decay of the translocation rates in Fig.~\ref{fig4}.

We now focus on the barrier-dominated regime $\sigma_m>\sigma_{m,2}$. Fig.~\ref{fig4} shows that the exponential decay of the 
translocation rate at $\sigma_m\approx\sigma_{m,2}$ is accompanied with the reversal of the polymer velocity. 
Indeed, in this regime with $\lb>\lel$, the capture velocity is negative, $v_{\rm dr}-v_b<0$, and one also gets from Eq.~(\ref{29I})
\be
\label{as2}
R_c\approx D\left(\lb-\lel\right)e^{-\lm\left(\lb-\lel\right)}.
\ee
The limiting law Eq.~(\ref{as2}) corresponds to the Kramers' transition rate formula associated with the electrostatic barrier 
$\Delta U\sim k_BT\lm(\lb-\lel)$ that has to be overcome by the polymer in order to penetrate the pore. 
Using Eqs.~(\ref{27II})-(\ref{lin1}) and~(\ref{lin2}), Eq.~(\ref{as2}) becomes
\bea
\label{as2II}
R_c&\approx&\frac{4\pi\ell_B}{g\kappa_b}\left[\frac{\ln(d/a)}{\eta\beta\kappa_bL_p}\sigma_p\sigma_m-\mu_eE(f_p\sigma_p-f_m\sigma_m)\right]\\
&&\times\exp\left\{-\frac{12\pi^2\ell_B\lm}{g\kappa_b\ln(L_p/2a)}\left[\frac{\ln(d/a)}{\kappa_b}\sigma_p\sigma_m\right.\right.\nonumber\\
&&\hspace{3cm}\left.\left.-\eta\beta L_p\mu_eE(f_p\sigma_p-f_m\sigma_m)\right]\right\},\nonumber
\eea
Eq.~(\ref{as2II})  explains the exponential decay of the translocation rates with $\sigma_m$ in the barrier-driven regime of Fig.~\ref{fig4}.

The threshold membrane charge $\sigma_{m,2}$ can be obtained from the equality $v_b=v_e$ together with Eqs.~(\ref{lin1}) and~(\ref{lin2}) as
\be\label{sigm2}
\frac{\sigma_{m,2}}{\sigma_p}= {f_p} \left[{f_m+\frac{\ln(d/a)\sigma_p}{\eta\beta\kappa_bL_p\mu_eE}} \right]^{-1}.
\ee
Comparison of Eqs.~(\ref{sigm1}) and~(\ref{sigm2}) shows that the characteristic charges for drift velocity inversion and transition from drift to 
barrier-driven regime satisfy $\sigma_{m,1}>\sigma_{m,2}$ (see also the inset of Fig.~\ref{fig2}). Thus, at membrane charges 
$\sigma_m \approx \sigma_{m,1}$, where the reversal of the drift velocity should occur, successful DNA capture events should be rare. 
This contradicts the suggestion of earlier works to reduce the polymer translocation velocity via the drift velocity inversion 
illustrated in Fig.~\ref{fig2}~\cite{the3}. We finally note that in Fig.~\ref{fig4}, the drift-dominated regime of longer polymers 
extends over an extended range of the membrane charge. Indeed, Eq.~(\ref{sigm2}) predicts that the rejection of longer 
polymers should occur at higher membrane charges, i.e. $L_p\uparrow\sigma_{m,2}\uparrow$. 
The mechanism behind this effect is investigated in Sec.~\ref{polyl}.

\begin{figure}
\includegraphics[width=1.\linewidth]{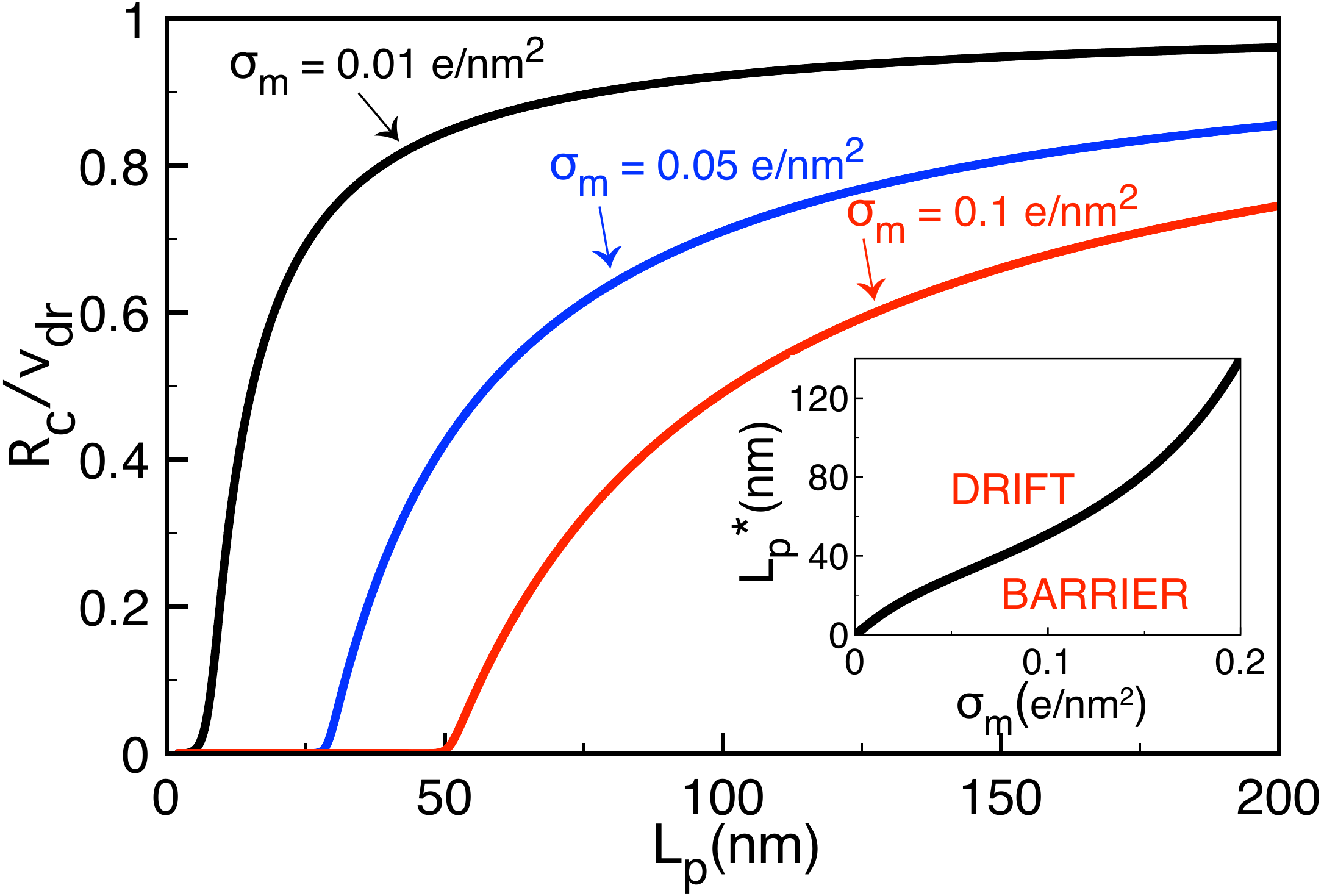}
\caption{(Color online) Main plot: Translocation rate $R_c$ rescaled by the drift velocity $v_{\rm dr}$ versus the polymer length 
$L_p$ at various membrane  charges. Inset: Threshold polymer length $L_p^*$ of Eq.~(\ref{Lp}) where the translocation rate 
becomes exponentially small versus the membrane charge $\sigma_m$. The model parameters are the same as in Fig.~\ref{fig2}.}
\label{fig5}
\end{figure}

Finally, in the inset of Fig.~\ref{fig4}, we display the behaviour of the translocation rate with the pore size. 
Beyond a characteristic pore size where one gets into the drift-dominated regime $\lel>\lb$, the translocation rate increases 
($d\uparrow R_c\uparrow$) and converges to the drift velocity $v_{\rm dr}$. This trend can be explained by the relation 
$R_c\approx v_{\rm dr}-v_b$ in Eq.~(\ref{as1}). The increase of the pore size reduces the membrane-induced potential 
$\phi_m(a)$ and the barrier $V_p(z_p)$. This lowers in turn the barrier-induced velocity component 
$v_b$ and the translocation becomes essentially drift-dominated at large pores, i.e. $R_c\approx v_{\rm dr}$. 
Next, we investigate the dependence of the translocation rates on the polymer length and voltage.

\subsubsection{Polymer length $L_p$ and voltage $\Delta V$}
\label{polyl}

In Fig.~\ref{fig5}, we display the behaviour of the rescaled translocation rate $R_c/v_{\rm dr}$ with the polymer length $L_p$. 
In qualitative agreement with experimental curves~\cite{e19,e22}, the translocation rate increases 
with the polymer length ($L_p\uparrow R_c\uparrow$) and saturates at the drift velocity $v_{\rm dr}$. 
This trend can be explained by Eq.~(\ref{as1II}) where the barrier-induced term decays as $L_p^{-1}$ while the drift term does not depend on 
$L_p$. The physical mechanism behind this peculiarity is encoded in the force balance Eq.~(\ref{a6}). 
One sees that the electric field $E$ acts on the whole polymer with length $L_p$ whereas the barrier-induced force 
$-V'_p(z_p)$ is induced exclusively by the polymer portion $l_p$ located in the pore. Hence, the longer the polymer, 
the stronger the drift effect with respect to the electrostatic barrier. This mechanism also explains the increase of the 
critical membrane charge $\sigma_{m,2}$ with the polymer length in Fig.~\ref{fig4}. 

Figure \ref{fig5} shows that due to the same mechanism, the stronger the membrane charge, the longer the characteristic polymer 
length $L_p^*$ where the translocation rate becomes vanishingly small, i.e. $\sigma_m\uparrow L_p^*\uparrow$. 
The length $L_p^*$ corresponding to the boundary between the barrier and drift dominated regimes follows from $\lel=\lb$ as
\be
\label{Lp}
L_p^*=-\frac{\ln(d/a)}{\eta\beta\mu_eE}\frac{a\sigma_p\phi_m(a)}{\delta\phi(d)-\delta\phi(a)}.
\ee
Equation (\ref{Lp}) is plotted versus the membrane charge in the inset of Fig.~\ref{fig5}. $L_p^*$ rises steadily with the membrane 
charge and its slope is amplified for $\sigma_m\gtrsim0.15$ $\mbox{e/nm}^2$. For the sake of analytical clarity, we pass to the linear 
PB approximation and expand Eq.~(\ref{Lp}) in terms of the charges $\sigma_m$ and $\sigma_p$. The critical polymer length simplifies to
\be
\label{Lplin}
L_p^*=\frac{\ln(d/a)}{\eta\beta\kappa_b\mu_eE}\frac{\sigma_m\sigma_p}{f_p\sigma_p-f_m\sigma_m}.
\ee
Equation (\ref{Lplin}) indeed predicts the increase of the critical length $L_p^*$ with the membrane charge for 
$\sigma_m<\sigma_{m,1}$ and its divergence at $\sigma_m\to\sigma_{m,1}$. This divergence reflects the fact that due 
to the reversal of the drift velocity at $\sigma_m=\sigma_{m,1}$, the drift effect cannot overcome the electrostatic barrier and 
drive the polymer into the pore regardless of how long the polymer is.

\begin{figure}
\includegraphics[width=1.\linewidth]{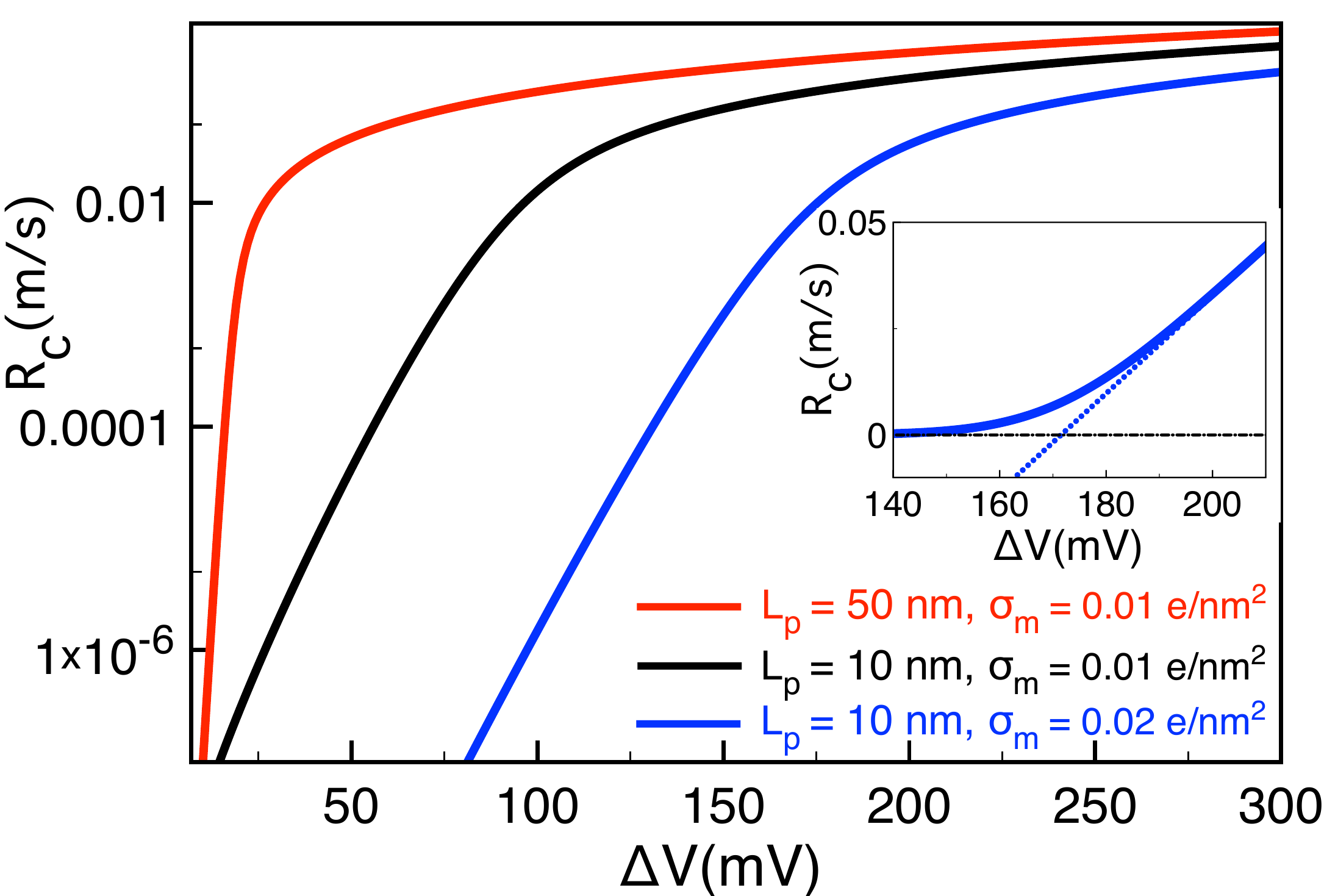}
\caption{(Color online)  Translocation rate $R_c$ versus voltage $\Delta V$ at various polymer lengths and membrane charges. 
The model parameters are the same as in Fig.~\ref{fig2}. The inset displays the translocation rate (solid curve) and the 
polymer capture velocity $v_{\rm dr}-v_b$ (dashed curve) at a linear scale.}
\label{fig6}
\end{figure}

In Fig.~\ref{fig6}, we display the evolution of the translocation rate with the voltage $\Delta V$ at various polymer lengths and membrane charges. 
Below a threshold voltage $\Delta V_*$ in the barrier-dominated regime of Eq.~(\ref{as2II}), 
the translocation rate increases exponentially with the external voltage. 
The same trend is illustrated in the inset at the linear scale. Above the threshold voltage $\Delta V_*$ where one
gets into the drift-dominated regime of Eq.~(\ref{as1II}),  the capture velocity switches from negative to positive and the 
translocation rate increases linearly with voltage. This turnover is in agreement with experiments~\cite{e19,e22} and 
simulations~\cite{the1}. The threshold voltage $\Delta V_*$ follows from Eq.~(\ref{Lplin}) as
\be
\label{dV}
\Delta V_*=\frac{\ln(d/a)L_m}{\eta\beta\kappa_bL_p\mu_e}\frac{\sigma_m\sigma_p}{f_p\sigma_p-f_m\sigma_m}.
\ee
In agreement with Fig.~\ref{fig6}, Eq.~(\ref{dV}) predicts the rise of the threshold voltage by the membrane charge $\sigma_m\uparrow\Delta V_*\uparrow$ and its reduction by the polymer length $L_p\uparrow\Delta V_*\downarrow$. Next, we characterize the effect of the polymer charge on the competition between the drift and barrier effects.

\subsubsection{Polymer charge $\sigma_p$}

\begin{figure}
\includegraphics[width=1.\linewidth]{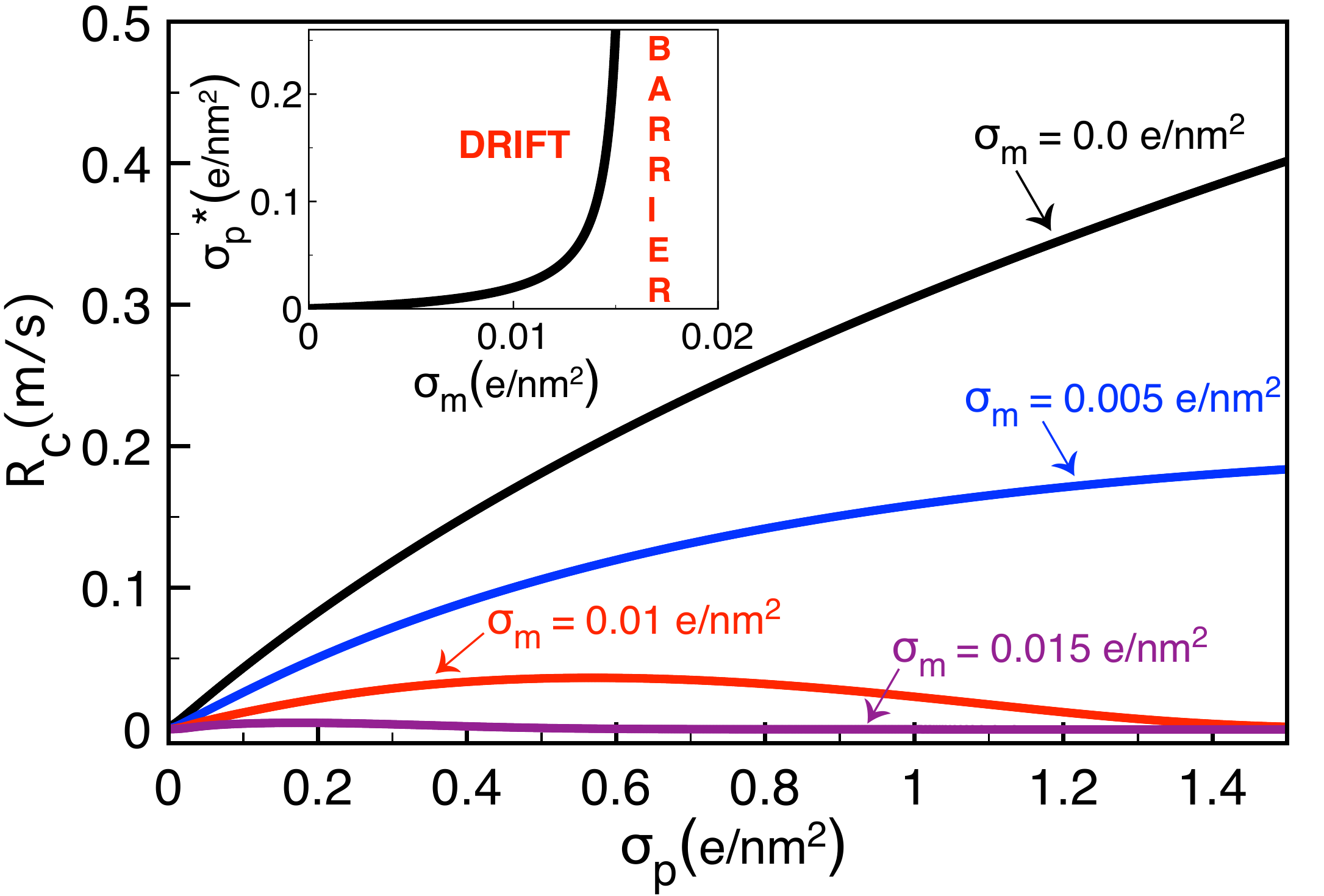}
\caption{(Color online) Translocation rate $R_c$ versus polymer charge density $\sigma_m$ at various membrane charges. 
The polymer length is $L_p=10$ nm. The remaining model parameters are the same as in Fig.~\ref{fig2}. 
The inset shows the critical polymer charge $\sigma_p^*$ of Eq.~(\ref{sigp}) where the translocation rate becomes vanishingly small.}
\label{fig7}
\end{figure}

The translocation rate Eq.~(\ref{as1II}) indicates that the opposing drift and barrier effects are both enhanced by the polymer charge 
$\sigma_p$. In order to understand the overall effect of the latter on the translocation process, in Fig.~\ref{fig7} we plot 
the translocation rate $R_c$ versus the polymer charge $\sigma_p$ at various membrane charges $\sigma_m$. 
In the case of a neutral pore $\sigma_m=0$ where translocation is driven by electrophoresis, due to the enhancement of the 
electrophoretic polymer mobility by the polymer charge the translocation rate increases monotonically. In charged pores 
where the electrostatic barrier component in Eq.~(\ref{as1II}) comes into play and reduces the amplitude of the translocation rate, 
the latter initially grows with the polymer charge ($\sigma_p\uparrow R_c\uparrow$), reaches a peak, and drops beyond this turning point ($\sigma_p\uparrow R_c\downarrow$) when the enhancement of the electrostatic barrier by the polymer charge takes 
over the amplification of the electrophoretic mobility. 

Figure \ref{fig7} shows that beyond the characteristic membrane charge $\sigma_m\approx0.015$ $\mbox{e/nm}^2$, 
regardless of the polymer charge strength, the translocation rate remains vanishingly small. In order to explain this peculiarity, we calculate the characteristic polymer charge $\sigma_p^*$ where the transition from the barrier to the drift-dominated regimes occurs. 
This follows by setting $R_c=0$ in Eq.~(\ref{as1II}),
\be
\label{sigp}
\sigma_p^*=\frac{f_m\sigma_m}{f_p\left[1-\sigma_m/\sigma_{m,3}\right]},
\ee
where we introduced the characteristic membrane charge
\be
\label{sigm3}
\sigma_{m,3}=\frac{f_p\eta\beta\kappa_bL_p\mu_eE}{\ln(d/a)}.
\ee
In the inset of Fig.~\ref{fig7} , the critical polymer charge Eq.~(\ref{sigp}) is seen to grow with the membrane charge and diverge 
at the threshold value $\sigma_{m,3}\approx0.016$ $\mbox{e/nm}^2$ beyond which translocation events become purely 
barrier-dominated at any polymer charge strength. The upper membrane charge $\sigma_{m,3}$ for successful translocation 
events is one of the key findings of our work. Eq.~(\ref{sigm3}) shows that this threshold charge increases with the polymer length 
$L_p$, the electric field $E$, and the salt density $\rho_b$. The effect of the salt density on the polymer translocation is thoroughly
 scrutinized in the next part.

\subsubsection{Salt concentration $\rho_b$}

\begin{figure}
\includegraphics[width=1.\linewidth]{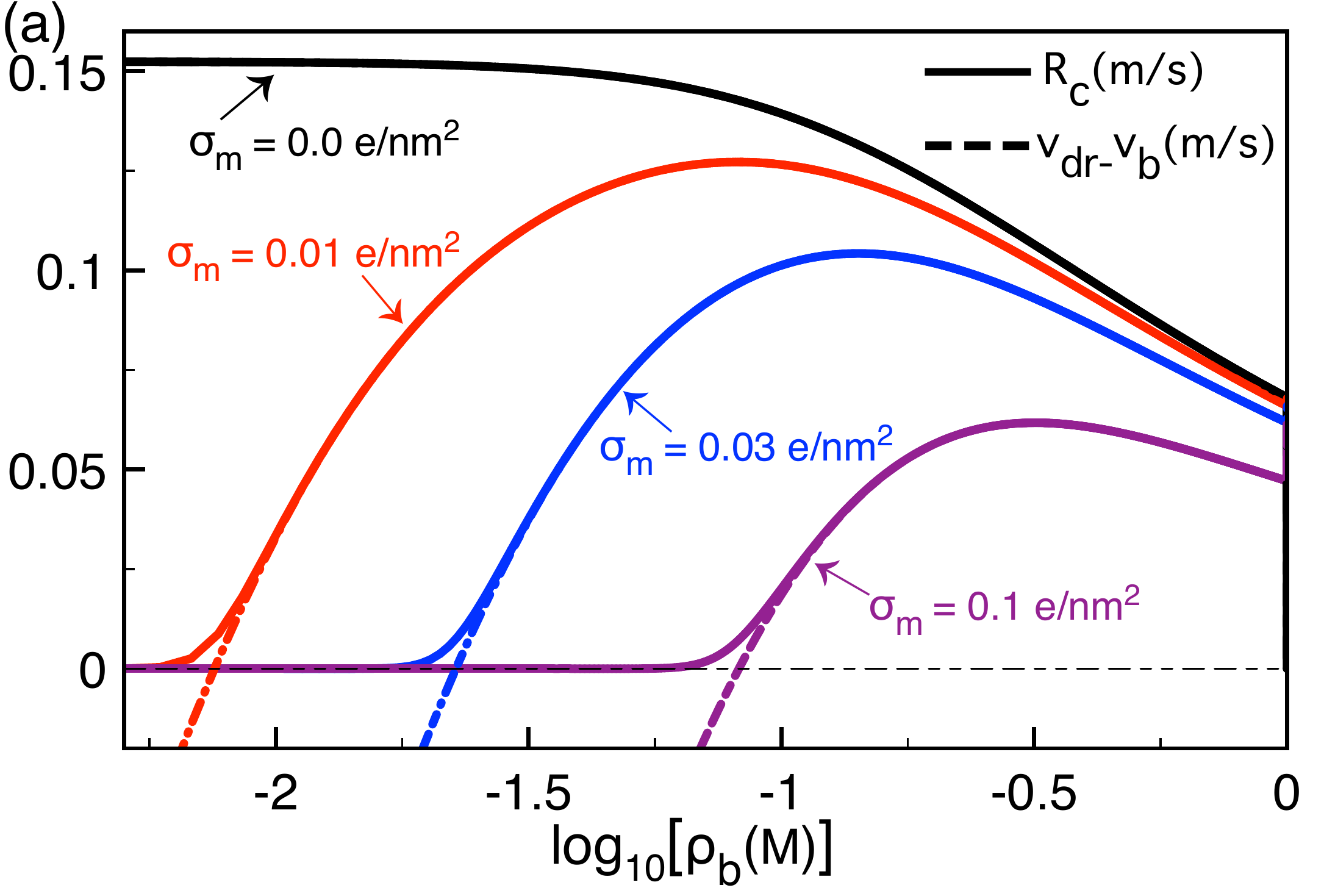}
\includegraphics[width=1.\linewidth]{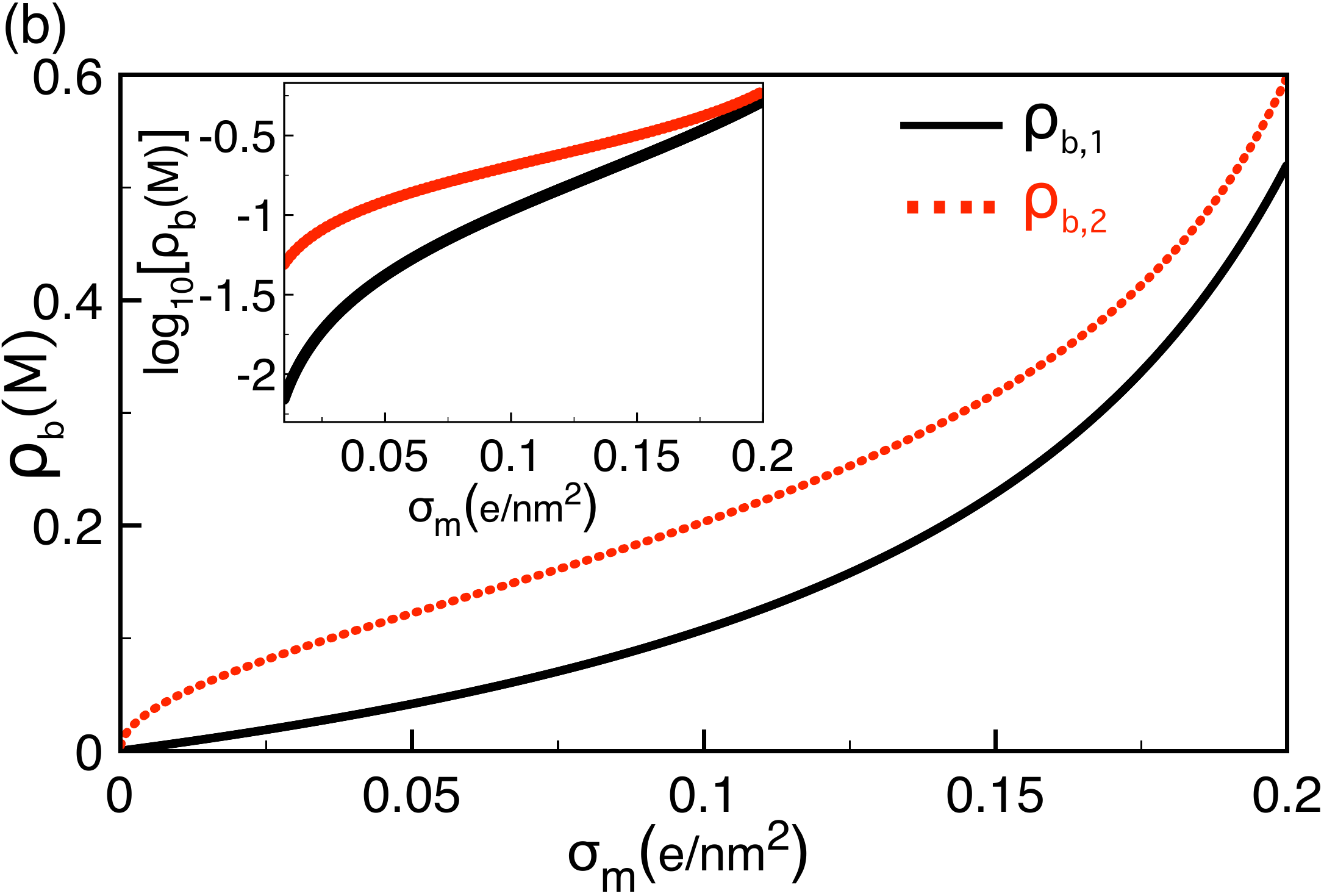}
\caption{(Color online) (a) Translocation rate (solid curves) and capture velocity (dashed curves) versus bulk salt 
density at various membrane charges. (b) Characteristic ion densities $\rho_{b,1}$ (Eq.~(\ref{crsalt1})) and $\rho_{b,2}$ 
(Eq.~(\ref{crsalt2})) against the membrane charge. The inset displays the main plot in a logarithmic scale. 
The model parameters are the same as in Fig.~\ref{fig2}.}
\label{fig8}
\end{figure}

Salt concentration is a practical control parameter that has not yet been fully considered in translocation experiments. 
This probably stems from our still incomplete understanding of the salt effects on the polymer capture and transport processes. 
Motivated by this point, in Fig.~\ref{fig8}(a), we illustrate the behaviour of  the translocation rates (solid curves) and capture 
velocities (dashed curves) with the salt density at various membrane charges. In order to interpret the curves, we 
Taylor expand Eq.~(\ref{as1II}) in terms of the screening parameter $\kappa_b$. This yields
\bea\label{salt}
R_c&\approx&4\pi\ell_B\mu_eE\left[a_p\sigma_p-a_m\sigma_m-8\pi\ell_B(b_p\sigma_p-b_m\sigma_m)\rho_b\right]\nonumber\\
&&-\frac{\sigma_p\sigma_m}{\eta\beta L_p}\frac{\ln(d/a)da}{\left(d^2-a^2\right)\rho_b},
\eea
where the auxiliary coefficients $a_{p,m}$ and $b_{p,m}$ that depend only on the pore and polymer radii are given in 
Appendix~\ref{ex}. In neutral pores $\sigma_m=0$ where the second term of Eq.~(\ref{salt}) associated with the barrier vanishes, 
the drift component of the translocation rate decreases linearly with the salt density $\rho_b$. 
In Fig~\ref{fig8}(a), the corresponding trend is shown by the black curve. This effect originates from the screening of the polymer 
charges and the resulting reduction of the electrophoretic polymer mobility.

In charged membranes, the barrier component of Eq.~(\ref{salt}) comes into play. In this case, Fig.~\ref{fig8}(a) shows that below a 
characteristic salt density $\rho_b=\rho_{b,1}$, translocation rates are vanishingly small. Beyond this salt density, due to the screening 
of the barrier component in Eq.~(\ref{salt}),  the translocation rates increase ($\rho_b\uparrow R_c\uparrow$), 
reach a maximum at $\rho_b=\rho_{b,2}$, and decrease in the purely drift-dominated regime ($\rho_b\uparrow R_c\downarrow$) 
where the charge screening of the polymer mobility occurs. The decreasing behaviour at strong salt concentrations was 
observed in translocation experiments where the increment of the salt density from $\rho_b=1$ M to $4$ M 
was shown to reduce the translocation rate by an order of magnitude~\cite{e22}.

The non-monotonic behaviour of the translocation rate with the salt concentration indicates that there exists an optimal 
concentration maximizing the probability of DNA capture into the pore. This result is one of the key predictions of our model.  
We first derive a closed form expression for the characteristic concentration $\rho_{b,1}$. In Eq.~(\ref{salt}), neglecting the first 
order correction coefficients $b_p$ and $b_m$,  and setting $R_c=0$, one gets
\be
\label{crsalt1}
\rho_{b,1}=\frac{\sigma_p\sigma_m}{4\pi\ell_B\eta\beta\mu_eEL_p(a_p\sigma_p-a_m\sigma_m)}\frac{\ln(d/a)da}{(d^2-a^2)}.
\ee
We calculate now the second characteristic salt concentration $\rho_{b,2}$ corresponding to the maximum of the curves in 
Fig.~\ref{fig8}(a). From the equation $\partial_{\kappa_b}R_c=0$, one finds
\be
\label{crsalt2}
\rho_{b,2}=\left[ \frac{\sigma_p\sigma_m}{32\pi^2\ell_B^2\eta\beta\mu_eEL_p(b_p\sigma_p-b_m\sigma_m)}\frac{\ln(d/a)da}{(d^2-a^2)}\right]^{1/2}.
\ee
Equations (\ref{crsalt1}) and~(\ref{crsalt2}) are plotted together in Fig.~\ref{fig8}(b). In agreement with the behaviour of the 
curves in Fig.~(\ref{fig8})(a), the characteristic ion concentrations $\rho_{b,1}$ and $\rho_{b,2}$  increase monotonically with the 
membrane charge density, i.e. $\sigma_m\uparrow\rho_{b,\{1,2\}}\uparrow$. Equation (\ref{crsalt2}) also shows that due to the 
amplification of the drift effect with respect to the electrostatic barrier, the larger the electric field or the longer the polymer, the lower the optimal salt concentration, i.e. $E\uparrow\rho_{b,2}\downarrow$ and $L_p\uparrow\rho_{b,2}\downarrow$. These predictions call for 
experimental verifications. We consider next the influence of the tunable experimental parameters on the polymer translocation time.

\subsection{Polymer translocation time}

In order to improve the accuracy of nanopore-based sequencing methods, one of the main challenges consists of 
adjusting the duration of the ionic current blockage induced by the translocating polymer. This objective clearly necessitates 
a high degree of control over the polymer translocation time. Motivated by this point, we 
characterize here the alteration of the polymer translocation time by tunable system parameters 
such as the pore charge and radius, and the bulk salt concentration. 

The translocation time corresponds to the mean first passage time of the polymer from the pore entrance at $z_p=0$ to the final point 
$z_p=L_m+L_p$ where the polymer leaves the pore. Substituting the current Eq.~(\ref{a2}) into the continuity 
Eq.~(\ref{a1}) and using the definition of the effective potential in Eq.~(\ref{a10}), the Smoluchowski equation takes the form of an 
effective Fokker-Planck equation
\be
\label{34}
\partial_tc(z_p,t)=D\partial_{z_p}^2c(z_p,t)+\beta D\partial_{z_p}\left[c(z_p,t)U'_p(z_p)\right].
\ee
In a stochastic process characterized by Eq.~(\ref{34}), the mean first passage time $\tau(z_2;z_1)$ from the initial point 
$z_1$ to the final point $z_2$ in the pore is given by the solution of the Dynkin equation~\cite{Risken},
\be
\label{35}
D\partial_{z_1}^2\tau(z_2;z_1)-\beta DU'_p(z_1)\partial_{z_p}\tau(z_2;z_1)=-1.
\ee
Solving Eq.~(\ref{35}) with reflecting and absorbing boundary conditions respectively at $z_1$ and $z_2$, one finds
\be
\label{36}
\tau(z_2;z_1)=\frac{1}{D}\int_{z_1}^{z_2}\mathrm{d}z'e^{\beta U_p(z')}\int_{0}^{z'}\mathrm{d}z''e^{-\beta U_p(z'')}.
\ee
Finally, we set $z_1=0$ and $z_2=L_p+L_m$ and carry out the double integral in Eq.~(\ref{36}) with the effective 
potential~(\ref{26}). After some algebra, one gets the translocation time $\tau\equiv\tau(0,L_p+L_m)$ in the form
\be\label{37}
\tau=\tau_1+\tau_2+\tau_3,
\ee
where the characteristic times for polymer capture, translocation, and exit are respectively given by
\begin{widetext}
\bea
\label{38I}
\tau_1&=&\frac{1}{D(\lel-\lb)^2}\left[e^{-(\lel-\lb)\lm}-1+(\lel-\lb)\lm\right];\\
\label{38II}
\tau_2&=&\frac{1}{D\lel(\lel-\lb)}\left[1-e^{-(\lel-\lb)\lm}\right]\left[1-e^{-\lel(\lp-\lm)}\right]+\frac{1}{D\lel^2}\left[e^{-\lel(\lp-\lm)}-1+\lel(\lp-\lm)\right];\\
\label{38III}
\tau_3&=&\frac{1}{D(\lel+\lb)^2}\left[e^{-(\lel+\lb)\lm}-1+(\lel+\lb)\lm\right]\\
&&+\frac{e^{-\lel(\lp-\lm)}}{D(\lel+\lb)}\left[1-e^{-(\lel+\lb)\lm}\right]\left\{\frac{1}{\lel-\lb}\left[1-e^{-(\lel-\lb)\lm}\right]+\frac{1}{\lel}\left[e^{\lel(\lp-\lm)}-1\right]\right\}.\nonumber
\eea
\end{widetext}

We consider now the simplest asymptotic limits of Eq.~(\ref{37}). In the limit of a vanishing electric field and neutral 
pore where $\lel=\lb=0$, the characteristic times~(\ref{38I})-(\ref{38III}) are purely diffusive and the capture time becomes 
$\tau_1=\lm^2/(2D)$. Then, the characteristic time associated with polymer penetration and translocation at constant length reads 
$\tau_1+\tau_2=\lp^2/(2D)$. Finally, the total translocation time becomes
\be
\label{39}
\tau=\frac{\left(L_m+L_p\right)^2}{2D}.
\ee
Thus, in the diffusive limit the translocation time increases quadratically with the polymer length $L_p$, which is a well-known result
for rodlike chains \cite{Tapsarev}. In the case of finite voltage $\Delta V$ and neutral pores, 
where the electrostatic barrier vanishes ($\lb=0$), the translocation time~(\ref{37}) takes the form
\be
\label{40}
\tau=\frac{(L_m+L_p)\lel-1+e^{-(L_m+L_p)\lel}}{D\lel^2}\approx\frac{L_m+L_p}{D\lel},
\ee
which yields the relation $L_m+L_p\approx v_{\rm dr}\tau$ characterizing a purely drift-assisted translocation. 
Equation (\ref{40}) shows that in the pure drift regime, the translocation time grows linearly with the polymer length 
$L_p$ and decays linearly with the voltage $\Delta V$. In the next subsection we scrutinize the alteration of the polymer translocation times by
membrane charge strength and pore confinement.

\subsubsection{Membrane charge $\sigma_m$ and pore radius $d$}

\begin{figure}
\includegraphics[width=1.1\linewidth]{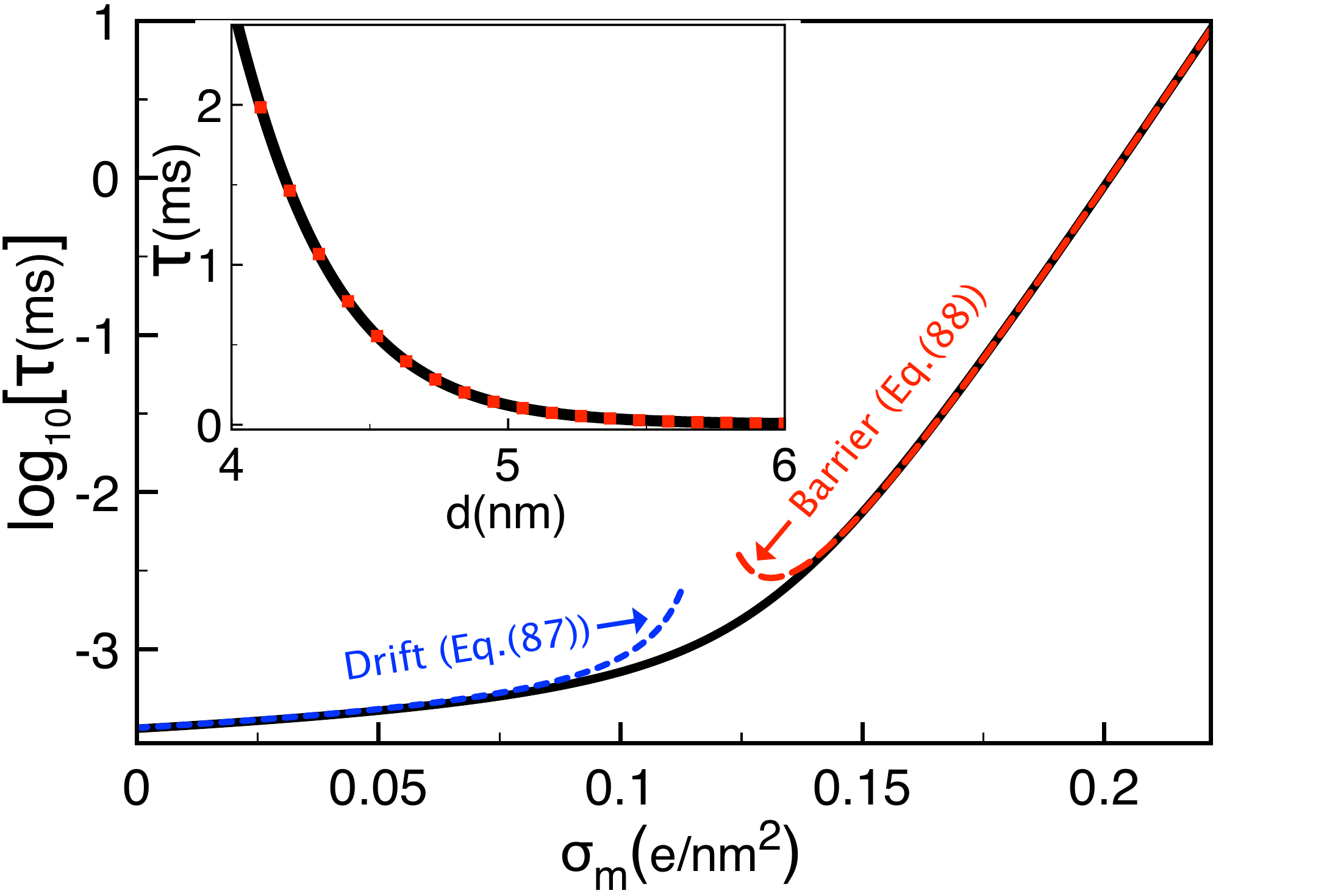}
\caption{(Color online) Translocation time Eq.~(\ref{37}) versus membrane charge (solid black curve). The dashed curves are the limiting laws 
with the corresponding equation numbers given in the legend. The inset displays the translocation rate versus pore size 
from Eq.~(\ref{37}) (solid curve) and its barrier limit of Eq.~(\ref{43}) (red symbols) at the membrane 
charge $\sigma_m=0.15$ $\mbox{e/nm}^2$. The salt concentration is $\rho_b=0.1$ M. 
The other model parameters are the same as in Fig.~\ref{fig2}.}
\label{fig9}
\end{figure}

The main plot of Fig.~\ref{fig9} displays the variation of the polymer translocation time~(\ref{37}) with the membrane charge 
density (solid black curve). In the region $\sigma_m<\sigma_{m,2}\approx0.12$ $\mbox{e/nm}^2$ corresponding to the drift-dominated 
regime, where the characteristic charge $\sigma_{m,2}$ is given by Eq.~(\ref{sigm2}), 
increasing the membrane charge weakly increases the translocation time. Beyond the membrane charge 
$\sigma_{m,2}$, where one switches to the barrier-driven regime, the translocation rate grows exponentially fast. 
More precisely, the alteration of the membrane charge by $\approx 0.1$ $\mbox{e/nm}^2$ enhances the translocation rate 
by four orders of magnitude. This strong sensitivity in the barrier-driven regime indicates that the chemical 
alteration of the membrane charge density can be an efficient way to tune the duration of ionic current signals in 
translocation experiments.  According to the black curves in Fig.~\ref{fig8}(b), the lower boundary 
$\sigma_{m,2}$ of this regime increases with bulk salt concentration, i.e. $\rho_b\uparrow\sigma_{m,2}\uparrow$. 
 
In order to understand the trend of the curves in Fig.~\ref{fig9}, one has to simplify Eqs.~(\ref{38I})-(\ref{38III}). 
Focusing on the experimentally relevant regime of strong electric fields $\lel L_{\pm}\gg1$ and neglecting exponentially small
terms, Eq.~(\ref{37}) simplifies as
\bea\label{41}
\tau&\approx&\frac{1}{D(\lel-\lb)^2}\left[e^{-(\lel-\lb)\lm}-1+(\lel-\lb)\lm\right]\nonumber\\
&&+\frac{1}{D\lel(\lel-\lb)}\left[1-e^{-(\lel-\lb)\lm}\right]+\frac{1}{D\lel(\lel+\lb)}\nonumber\\
&&+\frac{\lel(\lp-\lm)-1}{D\lel^2}+\frac{(\lel+\lb)\lm-1}{D(\lel+\lb)^2}.
\eea
In the drift-dominated regime $\lel>\lb$, neglecting the exponential terms of Eq.~(\ref{41}), one finds
\be
\label{42}
\tau\approx\frac{2\lel\lm}{D(\lel^2-\lb^2)}+\frac{\lp-\lm}{D\lel}.
\ee
For $\lel\gg\lb$, Eq.~(\ref{42}) tends to the pure drift limit of Eq.~(\ref{40}). Then, in the barrier-dominated regime $\lb>\lel$, 
by keeping only the exponential terms in Eq.~(\ref{41}), we get
\be
\label{43}
\tau\approx\frac{\lb}{D\lel(\lb-\lel)^2}e^{(\lb-\lel)\lm}.
\ee
Equations (\ref{42}) and~(\ref{43}) reported in Fig.~\ref{fig9} accurately reproduce the behaviour of the translocation time in the corresponding
regimes of validity. Taylor expanding the inverse distances in Eqs.~(\ref{25}) and~(\ref{25II}) in terms of the screening parameter $\kappa_b$, and the charge densities $\sigma_m$ and $\sigma_p$, we get
\bea
\label{l1}
\lel&\approx&\frac{3\pi\beta L_p eE}{\ln(L_p/2a)}\left[a_p\sigma_p-a_m\sigma_m-8\pi\ell_B(b_p\sigma_p-b_m\sigma_m)\rho_b\right];\nonumber\\
&&\\
\label{l2}
\lb&\approx&\frac{3\pi\ln(d/a)}{\ln(L_p/2a)}\frac{da}{d^2-a^2}\frac{\sigma_m\sigma_p}{\rho_b}.
\eea
One notes that the inverse lengths $\lel$ and $\lb$ scale linearly with the membrane charge $\sigma_m$. 
Considering this point, the asymptotic laws~(\ref{42}) and~(\ref{43}) explain the weak and the exponentially fast growth 
of the translocation time in the drift and barrier-dominated regimes of Fig.~\ref{fig9}, respectively. 
Finally, Eq.~(\ref{43}) indicates that in the barrier-driven regime, the translocation time decays exponentially with the external voltage. 
This agrees qualitatively with experiments and simulations~\cite{e14,the1}. 

In the inset of Fig.~\ref{fig9}, we display the variation of the polymer translocation time with the pore size. The exponential decay 
of the translocation time with the pore radius is in qualitative agreement with experiments on polymer transport through 
negatively charged silicon-based membrane nanopores (see Fig.7 of Ref.~\cite{e14}). The extension of the translocation 
time by a stronger confinement ($d\downarrow\tau\uparrow$) results from the amplification of the 
MF-level electrostatic barrier in the exponential of Eq.~(\ref{43}). Indeed, Eqs.~(\ref{43}) and~(\ref{l2}) show that with increasing pore size 
$d$, the translocation rate decays as $\ln\tau\sim 1/d$. We note in passing that due to the comparable range of the pore and 
polymer radii, the image-charge barrier neglected in our MF model is expected to enhance the total electrostatic barrier 
and the translocation time. This effect will be considered in a future work.

\subsubsection{Salt concentration $\rho_b$}

\begin{figure}
\includegraphics[width=1.1\linewidth]{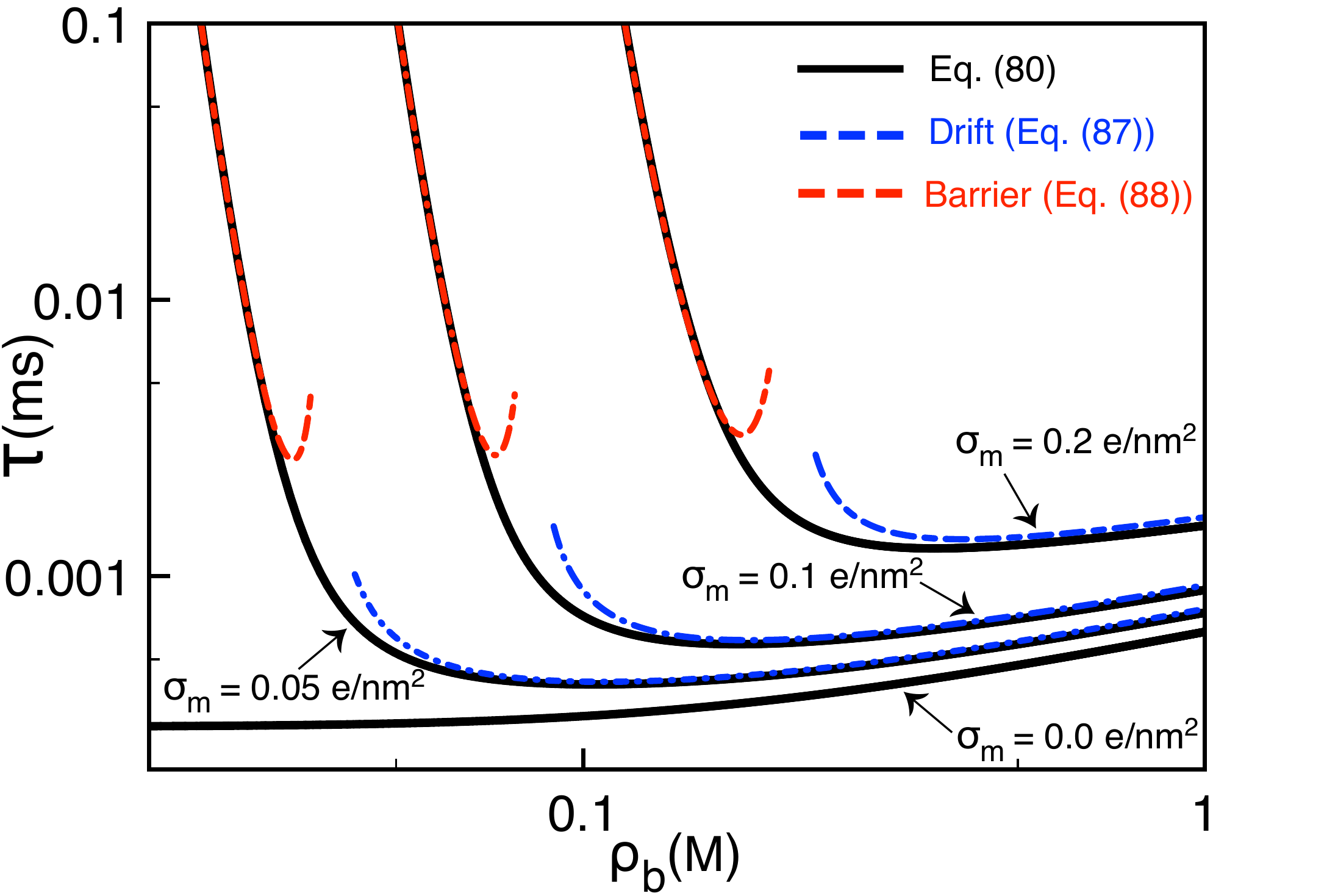}
\caption{(Color online) Translocation time Eq.~(\ref{37}) versus bulk salt concentration at various membrane charge 
densities (solid black curves). The dashed curves are the limiting laws with the corresponding equation numbers given in the legend. 
The model parameters are the same as in Fig.~\ref{fig2}.}
\label{fig10}
\end{figure}

In Fig.~\ref{fig10}, we display the salt dependence of the polymer translocation rate at various membrane charges (solid black curves). 
We also report the limiting laws of Eqs.~(\ref{42}) and~(\ref{43}) indicating the drift and barrier-driven regimes. In neutral pores 
where translocation is purely drift-driven, the increment of the salt density weakly affects the translocation time. 
In charged pores, due to the competition between salt screening of the electrostatic barrier and the electrophoretic DNA mobility, 
with increasing ion density, the translocation time drops in the barrier-dominated regime ($\rho_b\uparrow\tau\downarrow$), 
reaches a minimum, and weakly increases in the drift regime ($\rho_b\uparrow\tau\uparrow$). 

According to Eqs.~(\ref{43}) and~(\ref{l2}), in dilute salts the polymer translocation time decays with the ion density as 
$\ln\tau\sim1/\rho_b$ (see the red curves in Fig.~\ref{fig10}). This strong salt dependence of the translocation rate 
suggests that the alteration of the salt concentration in the barrier-driven regime can be an efficient way to tune the DNA 
velocity in translocation experiments. We finally note that in Fig.~\ref{fig10}, the minimum of the translocation time is located at the density 
$\rho_{b,2}$ given by Eq.~(\ref{crsalt2}). In agreement with the red curves in Fig.~\ref{fig8}(b), 
the increment of the membrane charge shifts the location of this minimum to larger salt concentration regimes, i.e. 
$\sigma_m\uparrow\rho_{b,2}\uparrow$.

\section{Summary and Conclusions}

Biopolymer translocation through nanopores under realistic experimental conditions remains a challenging problem due to the
complicated interplay between entropic, electrostatic and hydrodynamic degrees of freedom. In the present work we have focused
on the electrostatic interactions and
developed a consistent beyond-equilibrium theory of polymer capture and transport through charged pores in electrolyte solutions by coupling electrohydrodynamic equations with the Smoluchowski formalism. The main achievement from our theory is the incorporation of 
direct electrostatic polymer-membrane interactions to the polymer translocation velocity. In the relevant case of anionic polymers 
translocating electrophoretically through negatively charged pores, these interactions result in a repulsive electrostatic barrier 
$V_p(z_p)$ that reduces the polymer velocity from the drift value $v_{\rm dr}$ to $v_p(z_p)=v_{\rm dr}-\beta D^*V'_p(z_p)$. 
The corresponding competition between the electrostatic barrier and the drift effect gives rise to a critical membrane charge 
$\sigma_{m,3}=f_p\eta\beta\kappa_bL_p\mu_eE/\ln(d/a)$ above which  the polymer is likely to be rejected by the 
nanopore regardless of its charge strength (see Fig.~\ref{fig7}). The same competition results in a
non-monotonic behaviour of the polymer translocation rate with the bulk salt concentration (see Fig.~\ref{fig8}(a)). 
More precisely, due to the distinct ion density regimes where the salt screening of the electrostatic barrier and the 
electrophoretic polymer mobility occur, there exists a characteristic salt concentration 
$\rho_{b,2}$ given by Eq.~(\ref{crsalt2}) that maximizes the polymer capture probability. 
This prediction is of high degree of relevance to translocation assisted biopolymer sequencing. 

In addition, we investigated the influence of the electrostatic barrier on the polymer translocation time $\tau$. 
We found that in the barrier-dominated regime, the translocation time is highly sensitive to tunable system parameters. 
Namely, the translocation time rises exponentially fast with the membrane charge $\ln\tau\sim\sigma_m$, 
and decays exponentially with the pore size $\ln\tau\sim1/d$ and salt concentration $\ln\tau\sim1/\rho_b$. 
These features suggest that the variation of these parameters in the barrier-driven regime can be an 
efficient way to regulate the duration of the translocation process and the resulting ionic current blockage.

At this point we should highlight the approximations of our model and suggest potential improvements. First, in the computation of the 
membrane potential and the convective liquid velocity, we have neglected the edge effects associated with the finite thickness of the 
membrane. In order to relax this approximation that allowed us to keep the translational symmetry along the pore axis, one 
should account for the dependence of the electrostatic potential $\phi(\br)$ and convective velocity $u_c(\br)$  on the $z$ coordinate. 
This task can be achieved in the linear PB approximation where one should solve the linear PB and Stokes equations by the method 
of separation of variables. It should however be noted that this improvement will also increase the dimensionality of the problem 
and shadow the physical insight provided by our simpler theory. 

Second, we have treated electrostatic interactions at the MF level. This choice was motivated by the limitation of our work to 
monovalent electrolytes where correlations are known to play a minor role. However, in translocation experiments 
conducted with nanopores of size comparable with the polymer radius such as $\alpha$-Hemolysin pores, 
the strong confinement effects neglected by the MF electrostatics are expected to enhance the electrostatic 
barrier experienced by the polymer~\cite{e5}. In order to consider this complication as well as the effect of 
polyvalent salt on DNA transport where charge correlations are non-negligible, we plan to investigate electrostatic 
many-body effects in future work.

Finally, our polymer transport theory is based on a rigid polyelectrolyte model. Within the unified theory of charge and polymer 
fluctuations developed by Tsonchev {\it et al.}~\cite{dun}, the conformational fluctuations of the translocating polymer can 
be incorporated into the present transport theory in the future. Despite these approximations, our
various predictions have been shown to be in good qualitative agreement with translocation experiments and simulations. 
This indicates that our model embodies the most relevant features of these systems. We finally 
emphasize that our predictions on the effect of salt and membrane charge strength calls for experimental verifications.

\smallskip
\appendix
\section{Expansion coefficients}
\label{ex}

Here we list the expansion coefficients used in Eqs.~(\ref{crsalt1}) and~(\ref{crsalt2}):
\bea
\label{ex}
a_p&=&-\frac{a}{2}+\frac{ad^2\ln(d/a)}{d^2-a^2};\\
a_m&=&\frac{d}{2}-\frac{a^2d\ln(d/a)}{d^2-a^2};\\
b_p&=&-\frac{a}{32\left(d^2-a^2\right)^2}\left\{-\left(d^2-a^2\right)^2\left(3d^2-a^2\right)\right.\\
&&\left.\hspace{1.2cm}+4d^2a^2\ln(d/a)\left[4d^2\ln(d/a)-d^2+a^2\right]\right\}; \nonumber\\
b_m&=&-\frac{d}{32\left(d^2-a^2\right)^2}\left\{-\left(d^2-a^2\right)^2\left(d^2-3a^2\right)\right.\\
&&\left.\hspace{1.2cm}+4d^2a^2\ln(d/a)\left[-4a^2\ln(d/a)+d^2-a^2\right]\right\}. \nonumber
\eea


\begin{thebibliography}{99}
\bibitem {e1} J. J. Kasianowicz, E. Brandin, D. Branton, and D. W. Deamer, Proc. Natl. Acad. Sci. U.S.A \textbf{93}, 13770 (1996).  
\bibitem {e2} S. E. Henrickson, M. Misakian, B. Robertson, and J. J. Kasianowicz, Phys. Rev. Lett. \textbf{14}, 3057 (2000).  
\bibitem {e3} A. Meller, L. Nivon, and D. Branton, Phys. Rev. Lett. \textbf{86}, 3435 (2001). 
\bibitem {e4} A. Meller and D. Branton, Electrophoresis \textbf{23}, 2583 (2002).  
\bibitem {e5} D. J. Bonthuis, J. Zhang, B. Hornblower, J. Math\'{e}, B. I. Shklovskii, and A. Meller, Phys. Rev. Lett. \textbf{97}, 128104 (2006). 
\bibitem {e6} Y. Astier, O. Braha, and H. Bayley, J. Am. Chem. Soc. \textbf{128}, 1705 (2006). 
\bibitem {e7} D. Stoddart, A. J. Heron, E. Mikhailova, G. Maglia, and H. Bayley, Proc. Natl. Acad. Sci. U.S.A \textbf{106}, 7702 (2009). 
\bibitem {e8} J. Clarke, H.-C. Wu, L. Jayasinghe, A. Patel, S. Reid, and H. Bayley, Nature Nanotech. \textbf{4}, 265 (2009). 
\bibitem {e9} I. M. Derrington, T. Z. Butler, M. D. Collins, E. Manrao, M. Pavlenok, M. Niederweis, and J. H. Gundlach, Proc. Natl. Acad. Sci. U.S.A \textbf{107}, 16060 (2010). 
\bibitem {e10} H. Chang, F. Kosari, G. Andreadakis, M. A. Alam, G. Vasmatzis, and R. Bashir, Nano Lett. \textbf{4}, 1551 (2004). 
\bibitem {e11} P. Chen, J. Gu, E. Brandin, Y.-R. Kim, Q. Wang, and D. Branton, Nano Lett. \textbf{4}, 2293 (2004). 
\bibitem {e12} A. J. Storm, J. H. Chen, H. W. Zandbergen, and C. Dekker, Phys. Rev. E \textbf{71}, 051903 (2005).  
\bibitem {e13} R. M. M. Smeets, U. F. Keyser, D. Krapf, M.-Y. Wue, N. H. Dekker, and C. Dekker, Nano Lett. \textbf{6}, 89 (2006). 
\bibitem {e14} M. Wanunu, J. Sutin, B. Mcnally, A. Chow, and A. Meller,  Biophys. J. \textbf{95}, 4716 (2008). 
\bibitem {e15} R. F. Purnell, K. K. Mehta, J. J. Schmidt, Nano Lett. \textbf{8}, 3029 (2008). 
\bibitem {e16} H. Liu, J. He, J. Tang, H. Liu, P. Pang, D. Cao, P. Krstic, S. Joseph, S. Lindsay, and C. Nuckolls, Science \textbf{327}, 64 (2009). 
\bibitem {e17} M. Tsutsui, M.Taniguchi, K. Yokota, and T. Kawai, Nature Nanotech. \textbf{5}, 286 (2010). 
\bibitem {e18} M. Firnkes, D. Pedone, J. Knezevic, M. D\"{o}blinger, and U. Rant, Nano Lett. \textbf{10}, 2162 (2010). 
\bibitem{e19} M. Wanunu, W. Morrison, Y. Rabin, A.Y.  Grosberg, and A. Meller, Nature Nanotech. \textbf{5}, 160 (2010). 
\bibitem {e20} D. P. Hoogerheide, B. Lu, and J. A. Golovchenko, ACS Nano \textbf{8}, 7384 (2014). 
\bibitem{Tapsarev} V.V. Palyulin, T. Ala-Nissila, and R. Metzler, Soft Matter {\bf 10}, 9016 (2014).  
\bibitem {e21} S. Qiu, Y. Wang, B. Cao, Z. Guo, Y. Chen, and G. Yang, Soft Matter \textbf{11}, 4999 (2015). 
\bibitem {e22} N. A. W. Bell, M. Muthukumar, and U. F. Keyser, Phys. Rev. E \textbf{93}, 022401 (2016).  
\bibitem {the2} S. Ghosal, Phys. Rev. E \textbf{74}, 041901 (2006). 
\bibitem {the3} S. Ghosal, Phys. Rev. Lett. \textbf{98}, 238104 (2007). 
\bibitem {the4} J. Zhang and B. I. Shklovskii, Phys. Rev. E \textbf{75}, 021906 (2007).
\bibitem {the5} C.T.A. Wong and M. Muthukumar, J. Chem. Phys. \textbf{126}, 164903 (2007).
\bibitem {the7} M. Muthukumar, J. Chem. Phys. \textbf{132}, 195101 (2010).
\bibitem {the15} M. Muthukumar, J. Chem. Phys. \textbf{141}, 081104 (2014).
\bibitem {the8} A. Y. Grosberg and Y. Rabin, J. Chem. Phys. \textbf{133}, 165102 (2010). 
\bibitem {the12} P. Rowghanian and A. Y. Grosberg, Phys. Rev. E \textbf{87}, 042722 (2013). 
\bibitem {the13} P. Rowghanian and A. Y. Grosberg, Phys. Rev. E \textbf{87}, 042723 (2013). 
\bibitem {the14} P. Rowghanian and A. Y. Grosberg, J. Chem. Phys. \textbf{140}, 064701 (2014). 
\bibitem {the10} M. M. Hatlo, D. Panja, and R. van Roij, Phys. Rev. Lett. {\bf 107}, 068101 (2011).
\bibitem {the9} B. Luan and A. Aksimentiev, Soft Matter \textbf{6}, 243 (2010).  
\bibitem {the16} S. Buyukdagli and T. Ala-Nissila, Langmuir \textbf{30}, 12907 (2014).
\bibitem {Buyuk2017}  S. Buyukdagli, Phys. Rev. E \textbf{95}, 022502 (2017).
\bibitem {cyl1} Maria M. Tirado and J. Garc\'{i}a de la Torrea, J. Chem. Phys. \textbf{71}, 2581 (1979).
\bibitem {cyl2} A. Ortega and J. Garc\'{i}a de la Torrea, J. Chem. Phys. \textbf{119}, 9914 (2003).
\bibitem {math} M. Abramowitz and I.A. Stegun, \textit{Handbook of Mathematical Functions} (Dover Publications, New York, 1972).
\bibitem {the1} S. Matysiak, A. Montesi, M. Pasquali, A.B. Kolomeisky, and C. Clementi, Phys. Rev. Lett. \textbf{96}, 118103 (2006). 
\bibitem {Risken} H. Risken,\textit{The Fokker-Planck Equation} (Springer, Berlin, 1989).
\bibitem {dun} S. Tsonchev, R. D. Coalson, and A. Duncan, Phys. Rev. E \textbf{60}, 4257 (1999).
\end{thebibliography}
\end{document}